%% file: ms.tex
\documentclass[12pt,preprint]{aastex}

\newcommand{\aox}{$\alpha_{\rm ox}$}
\newcommand{\daox}{$\Delta \alpha_{\rm ox}$}
\newcommand{\aoxcorr}{$\alpha_{\rm ox}{\rm (corr)}$}
\newcommand{\daoxcorr}{$\Delta \alpha_{\rm ox}(corr)$}
\newcommand{\aoxl}{$\alpha_{\rm ox}(l_{2500})$}
\newcommand{\CIV}{\ion{C}{4}}

\newcommand{\MgII}{\ion{Mg}{2}}
\newcommand{\NV}{\ion{N}{5}}
\newcommand{\vmax}{$v_{\rm max}$}
\newcommand{\nh}{\mbox{${N}_{\rm H}$}}
\newcommand{\cmsq}{\mbox{\,cm$^{-2}$}}
\newcommand{\flux}{\mbox{\,erg\,cm$^{-2}$\,s$^{-1}$}}
\newcommand{\fnu}{\mbox{\,erg\,cm$^{-2}$\,s$^{-1}$\,Hz$^{-1}$}}
\newcommand{\flambda}{\mbox{\,erg\,cm$^{-2}$\,s$^{-1}$\,\AA$^{-1}$}}
\newcommand{\lumin}{\mbox{\,erg~s$^{-1}$}}
\newcommand{\kms}{\mbox{\,km\,s$^{-1}$}}
\newcommand{\xmm}{\emph{XMM-Newton}}
\newcommand{\chandra}{\emph{Chandra}}
\newcommand{\xspec}{\emph{XSPEC}}
\newcommand{\rosat}{\emph{ROSAT}}

\begin{document}
\title{The Correlation between X-ray and UV Properties of BAL QSOs}
\author{LuLu Fan,\altaffilmark{1,3}
HuiYuan Wang,\altaffilmark{1,2} Tinggui Wang,\altaffilmark{1,2}
Junxian Wang,\altaffilmark{1,2} Xiaobo Dong,\altaffilmark{1,2}
Kai Zhang,\altaffilmark{1,2} Fuzhen Cheng,\altaffilmark{1,2}}
\date{}

\altaffiltext{1}{Center for Astrophysics, University of Science and
Technology of China, Hefei, 230026, China;} \altaffiltext{2}{Joint
Institute of Galaxies and Cosmology, USTC and SHAO, CAS }
\altaffiltext{3}{SISSA/ISAS,Via Beirut 2-4,I-34014 Trieste,Italy}
\begin{abstract}
We compile a large sample of broad absorption lines (BAL) quasars
with X-ray observations from the \xmm\ archive data and Sloan
Digital Sky Survey Data Release 5. The sample consists of 41 BAL
QSOs. Among 26 BAL quasars detected in X-ray, spectral analysis is
possible for twelve objects. X-ray absorption is detected in all of
them. Complementary to that of \citet{gall06} (thereafter G06), our
sample spans wide ranges of both BALnicity Index (BI) and maximum
outflow velocity (\vmax\ ). Combining our sample with G06's, we find
very significant correlations between the intrinsic X-ray weakness
with both BALnicity Index (BI) and the maximum velocity of
absorption trough. We do not confirm the previous claimed
correlation between absorption column density and broad absorption
line parameters. We tentatively interpret this as that X-ray
absorption is necessary to the production of the BAL outflow, but
the properties of the outflow are largely determined by intrinsic
SED of the quasars.
\end{abstract}


\section{INTRODUCTION}

About 10\%-30\% of optically selected QSOs show broad absorption
lines (BAL) in their UV spectra, indicative of outflows with
velocities up to 0.1c\citep{hewett03,reichard03}. The similarity in
the UV continuum and emission lines between BAL and non-BAL QSOs
suggests that BAL QSOs are otherwise normal QSOs viewed in the
direction covered by the outflow \citep[e.g.][]{weymann91}. One
exception to these similarities is that BAL QSOs are soft X-ray
faint compared to non-BAL QSOs \citep[e.g.][]{green95,brinkmann99}.
The weakness in X-rays is interpreted as due to strong absorption
rather than intrinsic difference. Evidence for this has been
accumulated now from detailed studies of X-ray spectra of a few
bright BAL quasars, which display X-ray absorption with column
densities from $10^{22}$ to $\geq10^{24}$\cmsq\
\citep{wang99,gall99, gall02}.

Giving the ubiquity of X-ray absorption in BAL quasars, it is
natural to ask whether and how the X-ray absorbing gas is
connected to the UV BAL phenomenon. It has been known for quite long
time that BAL gas should be either confined into small clumps or
shielded from the intense soft X-rays in order to match the observed
profile. \citet{murray95} proposed that the highly ionized gas at
the base of disk wind (shielding gas) can naturally filter the soft
X-ray radiation to prevent the gas to be over-ionized so that the
radiative acceleration is effective \citep[See also][]{proga00}. As
both UV and X-ray absorbers are part of the continuous outflow, the
column densities of the two are expected to be correlated. Indeed,
\citet{brandt00} identified a correlation between the equivalent
width of \CIV\ absorption line and the soft X-ray weakness in a
sample of bright quasars, including half a dozen BAL QSOs.

\citet{wang05,wang07} found that electron scattering of the
shielding gas can explain the distribution of continuum polarization
in quasars, and the resonant scattering of BAL outflow can explain
the observed polarized spectrum of BAL. They further noted that
certain special features should appear in the polarized spectrum if
the size of the shielding gas is comparable with that of the BAL
outflows. As these features are only found in several low-ionization
BAL (LoBAL) QSOs  \citep[see their paper for details and
also][]{ogle99}, the shielding gas is likely well inside the BAL
outflow except in LoBAL QSOs. A similar conclusion has been reached
by studying the X-ray spectra of BAL QSOs
\citep[e.g.][]{gall04,gall06}. On the other hand, as
pointed by \citet{wang00}, in order to keep sufficient opacity in
the soft X-ray between $0.2-0.3$~keV, the absorber must have large
column density also of Li-like ions because those ions are responsible
for both the soft X-ray absorption between $0.2-0.3$~keV and the high
ionization UV BALs. Though this band is notoriously difficult to be studied,
they argued that at least in three bright low redshift BAL QSOs, the
X-ray absorption opacity around $0.2-0.3$ ~keV is large, suggesting
very large column density of Li-like ions. However, a relatively
small fraction of X-ray absorbing gas at moderate ionization level
will be sufficient to suppress the soft X-ray flux.

If the X-ray shielding is critical to the ionization balance in the
BAL outflow, which in turn affects the radiative accelerating force
on the outflow, one would expect that kinematic properties and
column density of BAL outflow will somehow correlate with the
properties of the X-ray absorber. In a sample of BAL quasars
observed by \chandra\ , G06 found a weak correlation between the
maximum outflow velocity(\vmax\ ) of BAL and the indicator (\daox\ )
of X-ray absorption. Their finding agrees with the qualitative
analysis that the strong soft X-ray absorption leads to more Li-like
ions, thus more efficiently radiative acceleration by UV photons.
However just as G06 pointed out that they had only four sources at
the low \vmax, and their sample of BAL QSOs is biased towards
strongly absorbed sources comparing to the BI distribution of SDSS
EDR BAL QSOs \citep{reichard03}. Giving the importance of this
question, more study based on a uniform sample is clearly required.

X-ray absorption is not the only factor that affects the ionization
equilibrium of BAL gas. \citet{steffen06} showed that the X-ray
luminosity of non-BAL QSOs has a large scatter for a given optical
luminosity. According to the current popular scenario that BAL and
non-BAL are only a matter of whether our line of sight passes
through BAL region or not, the intrinsic spectral energy
distribution (SED) between UV and X-ray of BAL QSOs should be also
diverse. Therefore, it would be interesting to study how the wind properties
depend on the intrinsic SED because ionization equilibrium is also
closely related to the intrinsic SED. If such a relation does exist,
it may offer insight into the driver of the outflows.

In this paper, we present a study of BAL QSOs from SDSS Data Release
5 (DR5) \citep{adelman07} that observed by \xmm\ satellite in X-ray
in order to explore the relations between UV and X-ray absorbers as
well as the relations between BAL properties and the intrinsic UV to
X-ray spectra. In \S2 we describe the selection of our \CIV\  BAL
QSOs sample and the data analysis in \S3. We show our results and
discuss the underlying physics in \S4. Finally, we summarize our
results in \S5. Throughout the paper, we assume the cosmological
parameters $H_0$ = 70 km s$^{-1}$ Mpc$^{-1}$, $\Omega_\mathrm{M}$ =
0.3 and $\Omega_{\Lambda}$ = 0.7.

\section{THE BAL QUASARS OBSERVED BY \xmm\ }\label{ref ss}

Starting from the spectroscopic quasar sample in the SDSS DR5
\citep{adelman07}, we compiled a sample of definitive \CIV\ BAL
quasars that have been observed by \xmm\ either serendipitously or
as a target. We restricted the redshift $1.5<z<4.0$ in order to make
sure that the \CIV\ $\lambda 1549$ is shifted into the SDSS
wavelength regime ($3800-9200$\AA). We matched these quasars with
\xmm\ pointing, and resulted in 225 quasars in the FOV of \xmm\
observation to date of April,2007.

We measured the BALnicity Index \citep[BI,][]{weymann91} and the
maximum outflow velocity for these 225 quasars using our own fitting
code (see \S3.2 for detail). We adopted the conventional definition
for the BAL QSO: the equivalent width (in \kms\ ) of any contiguous
absorption (at least $10\%$ below the continuum) exceeds 2000 \kms\
that falls between $3000-25000$ \kms\ blueshifted from the
systematic redshift \citep{weymann91}. All quasars with non-zero BI
were checked by eye, and ambiguous sources were removed.

Our final sample consists of 41 \CIV\ BAL quasars, including 5
LoBAL quasars, 22 HiBAL quasars, and 14 BAL quasars with unknown
BAL subtype because \MgII\ is not within the SDSS spctral coverage.
Most (25) of them have been included in the large BAL QSOs catalog
from SDSS DR3\citep{trump06}. In comparison with G06, BAL quasars
in our sample cover somewhat larger ranges of redshift ($1.579-3.776$)
and UV luminosity ($log(l_{2500})$:$30.212-32.230$), and are fainter
on average (31.055). Our sample has more uniform distributions in BI
($3-4610$ \kms\ with an average $1101$ \kms\ ) and in \vmax\
($5306-25000$ \kms\ with an average $14482$ \kms\ ) while G06's sample
consists mainly of BAL QSOs with the large BI (with an average $3437$
\kms\ ).

We notice that there are only four \xmm\ targeted objects: SDSS
J091127.61+055054.1, SDSS J111816.95+074558.1, SDSS J152553.89+513649.1
and SDSS J154359.44+535903.2 (Table 1). The first two are the lensed
BAL QSOs \citep{bade97}, which will be excluded from the following
correlation analysis. The third quasar was observed because of its
high optical polarization \citep{shemmer05}. The fourth object was
observed because of its X-ray detection by previous missions, thus
may bias towards X-ray bright sources \citep{grupe03}.

Some radio loud BAL QSOs show anomalous X-ray properties in
comparison with radio quiet counterparts
\citep{brotherton06,wang08}. In order to mark such sources, we
calculate the radio-to-optical flux ratios,
$R_{i}=log(S_{1.4GHz}/S_{i})$, following the definition of
\citet{ivezi02}.  The flux densities at 1.4~GHz, $S_{1.4GHz}$, are
taken from the Faint Images of the Radio Sky at Twenty centimeters
survey \citep[FIRST;][]{white97}. We estimate the $R_i$ upper limits
by taking the $2\sigma$ errors as the upper limit of radio flux
density. Only, three sources (SDSS J092345.19+512710.0,SDSS
J133004.72+472301.0 and SDSS J133553.61+514744.1) have radio
counterparts with the measured flux density of 1.72~mJy ,1.18~mJy
and 2.92~mJy, which give $R_i$=1.42, 1.18 and 1.14, respectively.

We listed our sample in Table 1 including the SDSS ID, the redshift,
$i$ band fiber magnitude of SDSS, the flux density at rest-frame
2500{\AA}($f_{2500}$),Galactic \nh\ from \citet{dickey90},the BAL
subtype and the radio-to-optical flux ratios $R_{i}$. Also we listed
the BI and \vmax\ in Table 1(see \S3.2). The values of $f_{2500}$
are calculated either by averaging the flux densities in the
rest-frame range of $2500\pm20$\AA\ or by extrapolating from the
continuum given by our fitting code.

\section{DATA ANALYSIS}

\subsection{X-ray Data Analysis}\label{sec xray}

The X-ray data were retrieved from \xmm\ Science Archive (XSA) and
prepared using $SAS\ 7.0.0$ with the most recent calibration files.
We extracted the background lightcurve above 10~keV, and the light
curve was used to filter the data obtained during the flaring
background periods using a threshold of 1.0 count~s$^{-1}$ for PN
and 0.5 count~s$^{-1}$ for MOS.
A sliding box cell detection algorithm ($eboxdetect$) was applied to
the images obtained by PN-CCD detector and two MOS-CCD detectors in
the soft ($0.3-2.0$~keV), hard ($2.0-10.0$~keV) and full
($0.3-10.0$~keV) bands to search for X-ray sources. We selected $L =
-ln(P)= 10$ as the minimum detection likelihood value, which in turn
corresponded to a probability of Poissonian random fluctuations of
the counts of $P = 4.5\times10^{-5}$. Among 41 BAL quasars, 26 were
detected in the full band  and 25 (13) were detected in the soft
(hard) bands at least on one EPIC instrument. The spectra were
accumulated from a circle region with a 30$\farcs$\ radius  except
for two sources locating close to the edge of the CCD whereas a
circle with 20$\farcs$\ radius was adopted. The backgrounds were
extracted from a source-free annulus surrounding each target on the
two MOS-CCD detectors and from a source-free circle along the
read-out direction on the pn-CCD detector. The photon counts were
extracted from the circular source regions centered on the SDSS
optical positions with the mentioned radius and the aperture
corrections were performed (See Table 2).
For non-detections the upper limits of counts are the $90\%$
confidence limits from Bayesian statistics\citep{kraft91}.
The redistribution matrix file ({\em rmf}) and
auxiliary response file ({\em arf}) were
generated using the tasks {\em rmfgen} and {\em arfgen}
respectively.

X-ray spectral modelling are performed using the package \xspec\
\citep{arnaud96}. We fit all the spectra with a uniform model, an
absorbed power law. It is found that the photon index $\Gamma$ of
the power-law is around 2.0 with a small scatter for radio-quiet
quasars \citep{george00,reeves00}. The broad band X-ray spectra of
BAL QSOs are quite similar to those of radio quiet non-BAL QSOs,
\citep{gall02,chart02,chartas03,aldcroft03,grupe03,page05},
therefore, in following analysis, $\Gamma$ is fixed to 2.0. Both the
Galactic neutral HI absorption and an intrinsic absorption are
included in the model. The Galactic neutral HI column density is
fixed at the value derived from Galactic HI maps\citep{dickey90}
(See Table 1). Due to limited count rates and the large
uncertainties, we do not consider more complex X-ray absorption
models (e.g., a partially covering or ionized absorber), and just
adopt a simple neutral absorption with a solar chemical composition
at the source rest frame ({\em zwabs} in \xspec\ )
\citep{morrison83} for the intrinsic absorption.

To deal with very different X-ray counts available, two different
methods are used to estimate the intrinsic absorption column
density.  The X-ray spectrum is fitted directly with an absorbed
power-law if the source is detected in the both soft and hard band.
Only the intrinsic absorption and the power-law normalization are
free parameters. Other parameters, including redshift, the Galactic
neutral absorption and the photon index $\Gamma$, are fixed to the
proper values. Either $\chi^2$-statistic or C-statistic
\citep{cash79} is taken as the merit of the fit depending on the net
source counts. If the net source counts are greater than 100, the
spectrum is re-binned with at least 15 counts per bin, and the fit
is performed by minimizing $\chi^2$. Otherwise, the spectrum is not
binned and the fit is performed by minimizing the C-statistics. In
order to test the validity of the fixed $\Gamma$ power-law model, we
make $\Gamma$ free to fit eight sources with greater than 100 counts
.The average value of $\Gamma$ of six sources is about 1.90, which
is very close to 2.0.The other two (SDSS J091127.61+055054.1 and
SDSS J100728.69+534326.7) show a rather flat spectra with $\Gamma
\sim 1.2$. \citet{page05} mentioned that SDSS J091127.61+055054.1
should be better modelled by a broken power law with the values of
$\Gamma$ 0.92 and 1.96,respectively.
The fitted intrinsic absorptions are very similar to those
listed in Table 3 except the two flat spectral sources. A flat spectrum
can be caused by complex absorption, strong reflection component or an
intrinsic flat power-law. Because all
other six sources show normal X-ray spectra and there is no evidence
for strong Fe\,K$\alpha$ in the X-ray spectrum, we believe that the
flat spectra in these two objects are caused by complex absorptions.
However, due to limit counts available, we will not try more
complicated models.

For those sources detected only in hard or soft band, the
upper/lower limit of the column density of intrinsic absorption is
estimated from the hardness ratios, defined as ${\rm HR} =
(h-s)/(h+s)$, where {\em h} and {\em s} are referred to the hard and
soft band counts, respectively. First for each source, we calculate
hardness ratios for a grid absorbed power-law models with column
densities in the range of $10^{20}-10^{24}$\cmsq\ using the {\em
arf} and {\em rmf} at the source position. The observed hardness
ratio is then compared to the models and then the upper/lower limit
of the intrinsic absorption could be derived.

In order to compare our sample with that of G06, we also calculate
\aox\ , \daox\ and \aoxcorr\ ,\daoxcorr\ , defined in the G06
\footnote{Our definition of the soft($s:0.3-2.0$~keV) and hard
($h:2.0-10.0$~keV) bands is slightly different from
theirs($s:0.5-2.0$~keV;$h:2.0-8.0$~keV).}, as follows. First, the
above hardness ratios for all BAL QSOs in the sample were estimated.
Hardness ratios for a grid of power-law models only absorbed by the
Galactic column density in that direction were then calculated using
the \xspec. The observed HR was compared to the model HRs to
estimate the photon index $\Gamma_{\rm HR}$. With the best fitted
photon-index, the normalization at 1~keV was derived from the
count-rates. The Galactic absorption corrected 2~keV flux is
determined from the model. We defined the UV to X-ray broad band
spectral index as \aox\ $=0.384log(f_{2keV}/f_{2500})$
\citep{tana79}. It is found that \aox\ is correlated with the
optical luminosity of quasars \citep{yuan98,avni86,wilkes94,green95},
though its reality has been questioned by \citet{yuan98,tang07}.
Following G06, we introduce a
quantity of \daox\ = \aox\ $ -$\aoxl\ to characterize the weakness
of the X-ray emission of the quasar relative to the average quasars
at that UV luminosity, where \aoxl\ was the expected \aox\ based on
the 2500 \AA\ monochromatic luminosity, $l_{2500}$\citep{stra05}.For
sources not detected in both band, an upper limit to the X-ray flux
was derived by assuming $\Gamma=1.0$. Finally, \aoxcorr\ and
\daoxcorr\ were calculated from an ``absorption-corrected" value for
the 2~keV flux density estimated by a fixed $\Gamma = 2.0$ power-law
model normalized by the counts rate in the observed-frame 2-10~keV
bandpass. Note that for two lensed sources, SDSS J091127.61+055054.1
and SDSS J111816.95+074558.1, we do not calculate \daox\ or
\daoxcorr\ since their intrinsic $l_{2500}$ are unknown.

\subsection{Ultraviolet Spectral Analysis And \CIV\ Absorption-Line
Parameters}

Following the procedures described in \citet{zhou06}, we calculate
both the BI of \CIV\ and \MgII\ absorption lines and their maximum
outflow velocity, \vmax. Briefly, we use the SDSS composite quasar
spectrum \citet{vanden01} as the template for continuum and emission
line spectrum.  The template is reddened and scaled to match the
observed quasar spectrum in the absorption line free windows. The BI
of \CIV\ and \MgII\ are calculated following the definition given by
\citet{weymann91} and \citet{reichard03}, respectively, as follows,
\begin{equation}
BI=\int^{25000}_{0\ or\ 3000}dv[1-\frac{F^{obs}(v)}{0.9F^{fit}(v)}]C(v)
\end{equation}
where $F^{obs}(v)$ and $F^{fit}(v)$ are the observed and fitted
fluxes, respectively, as a function of velocity in \kms\ from the
systematic redshift within the range of each absorption trough and
\begin{equation}
C(v)=\left\{
\begin{array}{ll}
1.0,&\quad \mbox{if $[1-\frac{F^{obs}(v)}{0.9F^{fit}(v)}]>0
\  over\ a\ continuous\ interval\ of\ \gtrsim W\  \kms\ $} \\
0,&\quad   \mbox{otherwise}
\end{array}
\right.
\end{equation}
The integral in equation (1) starts from $v=3000$ \kms\ for \CIV\
and from $v=0$ \kms\ for \MgII. The threshold interval in equation
(2) is $W=2000$ \kms\ for \CIV\ and from $W=1000$ \kms\ for \MgII.
Five sources show the non-zero BI of \MgII\ consistent with $\sim
10\%$ fraction of low ionization BAL QSOs. The maximum outflow
velocity are calculated, simultaneously. The BI and \vmax\ of \CIV\
absorption lines are listed in Table 1.

Finally, we also remeasure the BI and \vmax\ for G06's sample on the
LBQS spectra
\citep{foltz87,foltz89,hewett91,chaffee91,morris91,hewett95} using
our method because we will combine G06's sample with ours in the
statistical analysis between the properties of X-ray and UV. We find
that our measurements of either BI or \vmax\ are well correlated
with those of G06 although there is considerable scatter. For six
objects both in this sample, the SDSS BAL QSO sample of
\citet{trump06} and in G06, our measurements appear in between
theirs. Note that the results of correlation analysis by using G06's
BI and \vmax\ are very similar to those obtained by using ours.

\section{RESULTS AND DISCUSSION}\label{sec_res}

\subsection{X-ray Properties Of BAL QSOs}\label{sec xrayprop}

To investigate the general X-ray properties of our SDSS/\xmm\ BAL
sample (this paper), we try to measure the intrinsic absorption
adopting a simple neutral absorption. However, only 12 of 41 sources
can be fitted directly to give the intrinsic absorption column
densities \nh\ in the range $\sim 4\times10^{21}$ to $\sim
2\times10^{23}$\cmsq. For 14 sources, upper/lower limits can be
placed by the hardness ratios HR. The final sample spans a wide
range of intrinsic absorption column densities from $<10^{20}$\cmsq\
to $\sim 10^{24}$\cmsq\ (Table 3). The lowest limit is obtained for
the LoBAL QSO, SDSS J092238.43+512121.2. We have rechecked the
optical spectrum, the identification of this quasar as LoBAL might
be questionable because of the presence of narrow absorption lines.
Notably, five LoBAL QSOs do not show stronger absorption than HiBAL
QSOs.

\begin{figure}[t]
\plotone{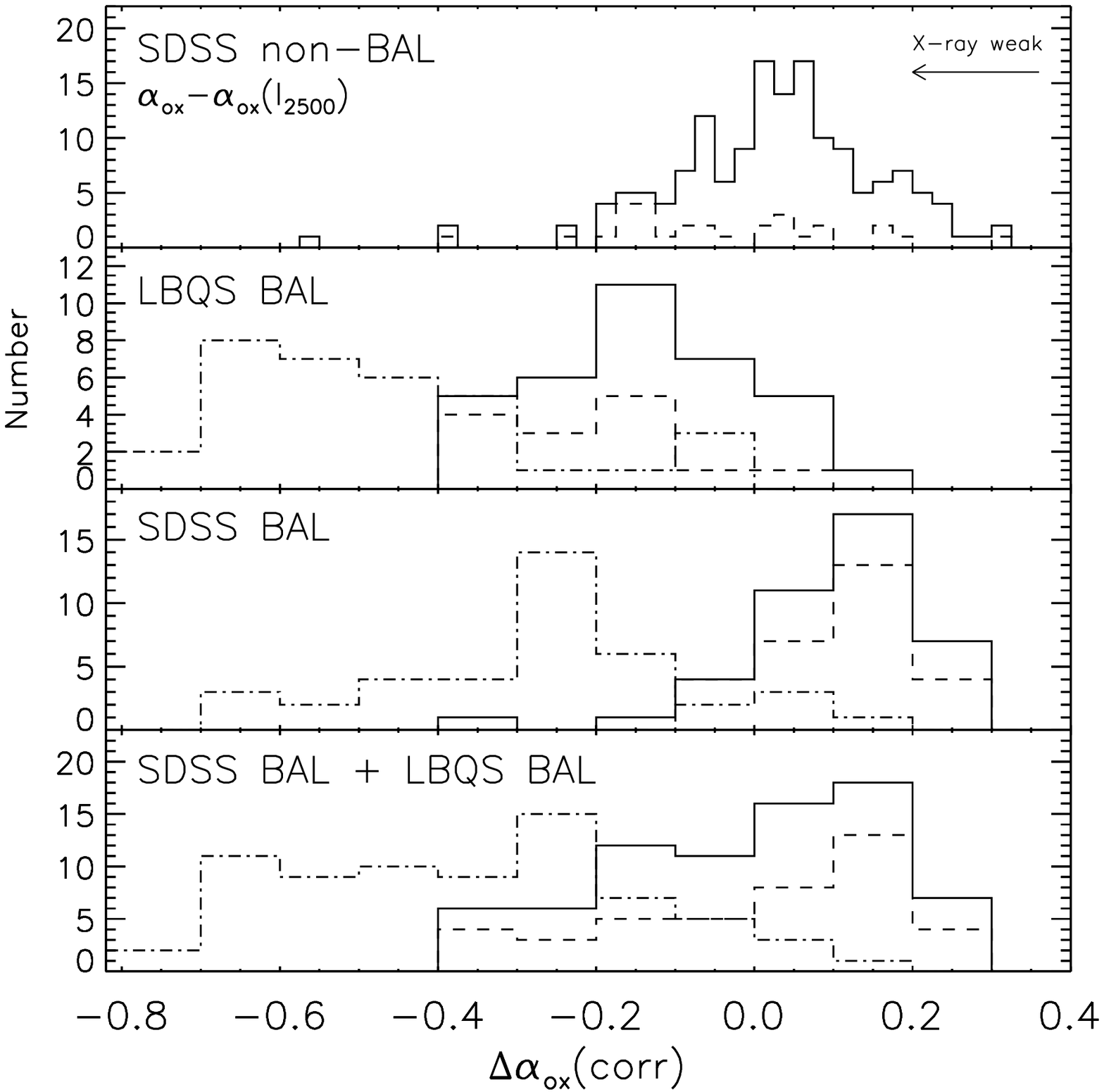}\caption{The top panel shows the distribution of
observed \daox\ = \aox\ $-$ \aoxl\ for the SDSS/\rosat\ non-BAL
sample\citep{stra05}.The three lower panels show the distributions
of \daoxcorr\ = \aoxcorr\ $-$ \aoxl\ for the LBQS/\chandra\ BAL
sample(G06),the SDSS/\xmm\ BAL sample and the combined sample(this
paper),respectively.For all four panels,solid lines indicate the
full samples and dashed lines only show upper limits.Dot-dashed line
in the three lower panels represents the distribution of \daox.The
arrow in the top panel shows the direction of the X-ray weak objects
for our convention of \aox. }\label{fig_xdis}
\end{figure}

Following G06,we measure \aox\ or place upper/lower limits on it,
which ranges from $-1.36$ to $-2.26$ with an average $-1.86$ (Table
3). Similar to G06, we show the \daox, to account for the luminosity
dependence of \aox(Table 3,see also the dot-dashed line in Fig.
\ref{fig_xdis}). The average value of \daox\ is $-0.25$, suggesting
that 2~keV X-ray luminosities (at rest frame) of our SDSS/\xmm\ BAL
sample are roughly three times fainter than the SDSS/\rosat\ non-BAL
sample\citep{stra05}. And in Table 3 we also present the \aoxcorr\
and \daoxcorr\ as a surrogate of the intrinsic X-ray properties of
BAL QSOs. \aoxcorr\ is calculated by assuming $\Gamma=2.0$ and using
the hard-band counts rate to normalize the X-ray continuum and
\daoxcorr\ =\aoxcorr\ $-$ \aoxl\ (see G06 or \S3.1 for the
definition). We find \daoxcorr\ in the range from $-0.36$ to $0.29$
with an average value of 0.11, which indicates our SDSS/\xmm\ BAL
sample is slightly X-ray brighter, relative to the average quasars
at that UV luminosity, than the SDSS/\rosat\ non-BAL
sample\citep[see Fig. \ref{fig_xdis}]{stra05}.The X-ray brighter of
our sample may be due to the relative shallower detection threshold
of \xmm\ relative to \chandra . Comparing the \daoxcorr\
distribution of the LBQS/\chandra\ BAL sample(G06) with the
SDSS/\rosat\ non-BAL sample\citep[figure 2 of G06]{stra05} one can
find the LBQS/\chandra\ BAL sample(G06) is slightly intrinsic X-ray
weaker with a median \daoxcorr$=-0.14$ than the normal QSOs.
Alternately, even hard X-rays are absorbed in the LBQS/\chandra\ BAL
sample(G06) so that the simple assumption is broken down (See G06 or
\S3.1 for detail). We have carried out simulations to test this
effect. Using \xspec, we simulate the dependence of the \daoxcorr\
on the varying neutral hydrogen column density \nh. We assume that
\daoxcorr\ equals to zero for a single $\Gamma=2$ power-law with
\nh$=10^{20}$\cmsq\ at the redshift 2. We find that an absorption
column density of $3\times10^{23}$\cmsq\ is required in order to
account for the mean offset, about $-0.14$, of \daoxcorr\ of the
LBQS/\chandra\ BAL sample(G06) relative to the SDSS/\rosat\ non-BAL
sample\citep{stra05}. If this is the main cause, most of X-ray weak
sources in the LBQS/\chandra\ BAL sample(G06) will have a column
density at least order of this. Future X-ray observation is
certainly needed to assess this.


Either intrinsic X-ray weak or large column density of absorber may
indicate that the LBQS/\chandra\ BAL sample(G06) is biased in X-ray
properties. Their sample obviously has larger values of BI and
\vmax\ than the SDSS BAL QSOs\citep{reichard03} and is not uniform
on UV properties too. Note that our SDSS/\xmm\ BAL sample is more
uniform, especially on UV properties, and can be used as a
complement to the LBQS/\chandra\ BAL sample(G06). For direct
comparison, we show the distribution of \daox\ for the SDSS/\rosat\
non-BAL sample\citep{stra05} and \daoxcorr\ for our SDSS/\xmm\ BAL
sample(this paper) and the LBQS/\chandra\ BAL QSO sample(G06) in
Fig. \ref{fig_xdis}. In the following we used the combined sample of
ours 41 and G06's 35 sources to study the relations between X-ray
and UV properties. Note again, the UV properties of the
LBQS/\chandra\ BAL QSO sample(G06) used in this paper are obtained
by using our procedures so that we can use the consistent definition
of BI and \vmax. We also use the hardness ratios presented in G06 to
calculate \nh\ of these G06 QSOs following the same approach as we
have done for \xmm\ sources.

\subsection{X-ray And UV Absorptions}\label{sec_ab}

One of the purposes of this paper is to study the relationship between
the UV and X-ray absorbers. It is generally believed that the
X-ray absorber shields the disk winds from soft X-rays and
makes line driving more efficient. A naive deduction is
that the properties of UV and X-ray absorptions are correlated.
Basing on this idea, G06 presented a correlation analysis between
the X-ray absorption using \daox\ as an indicator and the UV
absorption properties such as BI, DI, \vmax\ and $f_{\rm deep}$.
They found only a weak correlation between \daox\ and \vmax. We
will carry out a similar analysis using a larger sample covering
more uniformly the whole BI range. In the following analysis, we
will use Kendall-$\tau$ test to quantify the significance of a
correlation.

\begin{figure}[t]
\plottwo{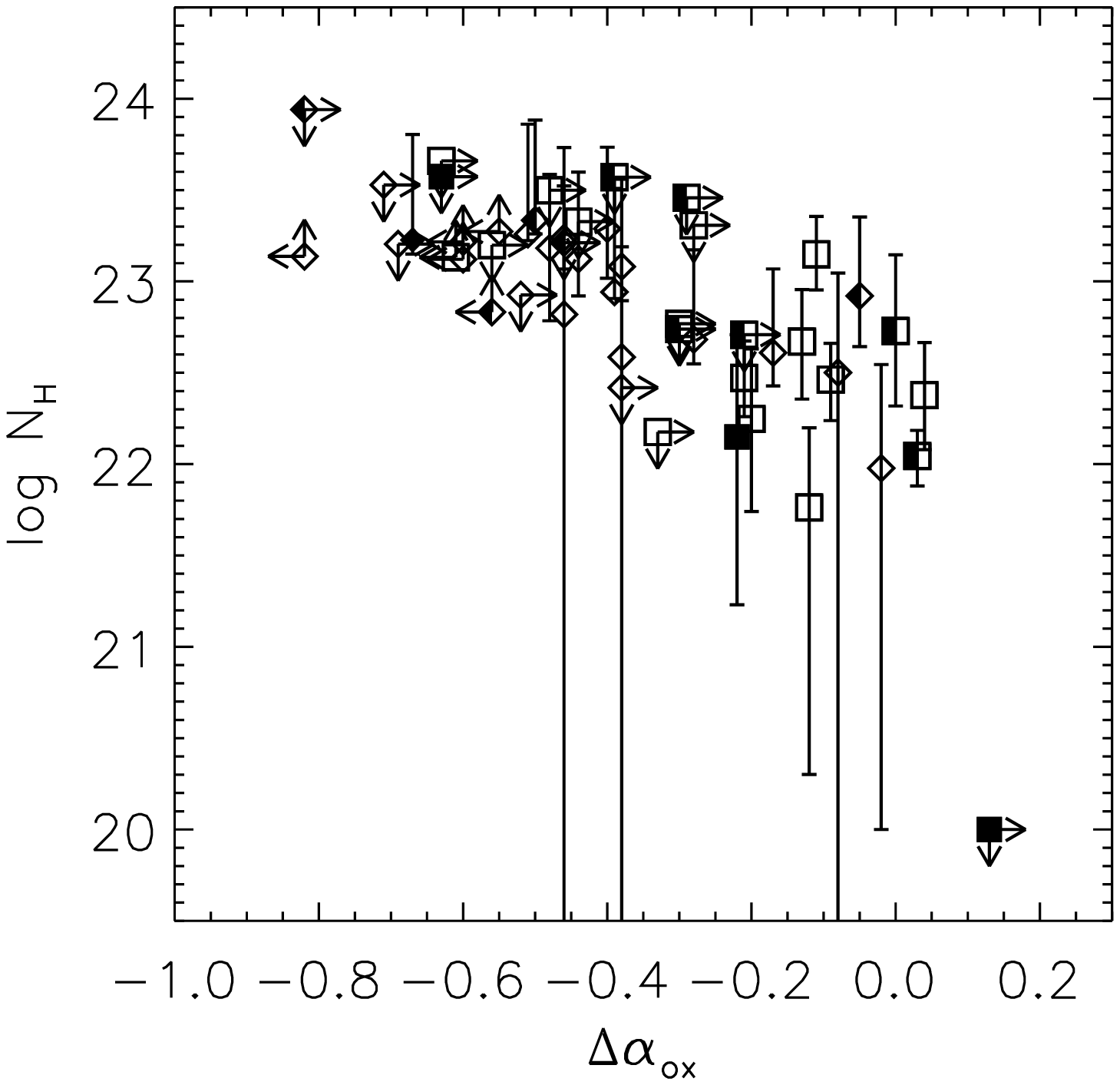}{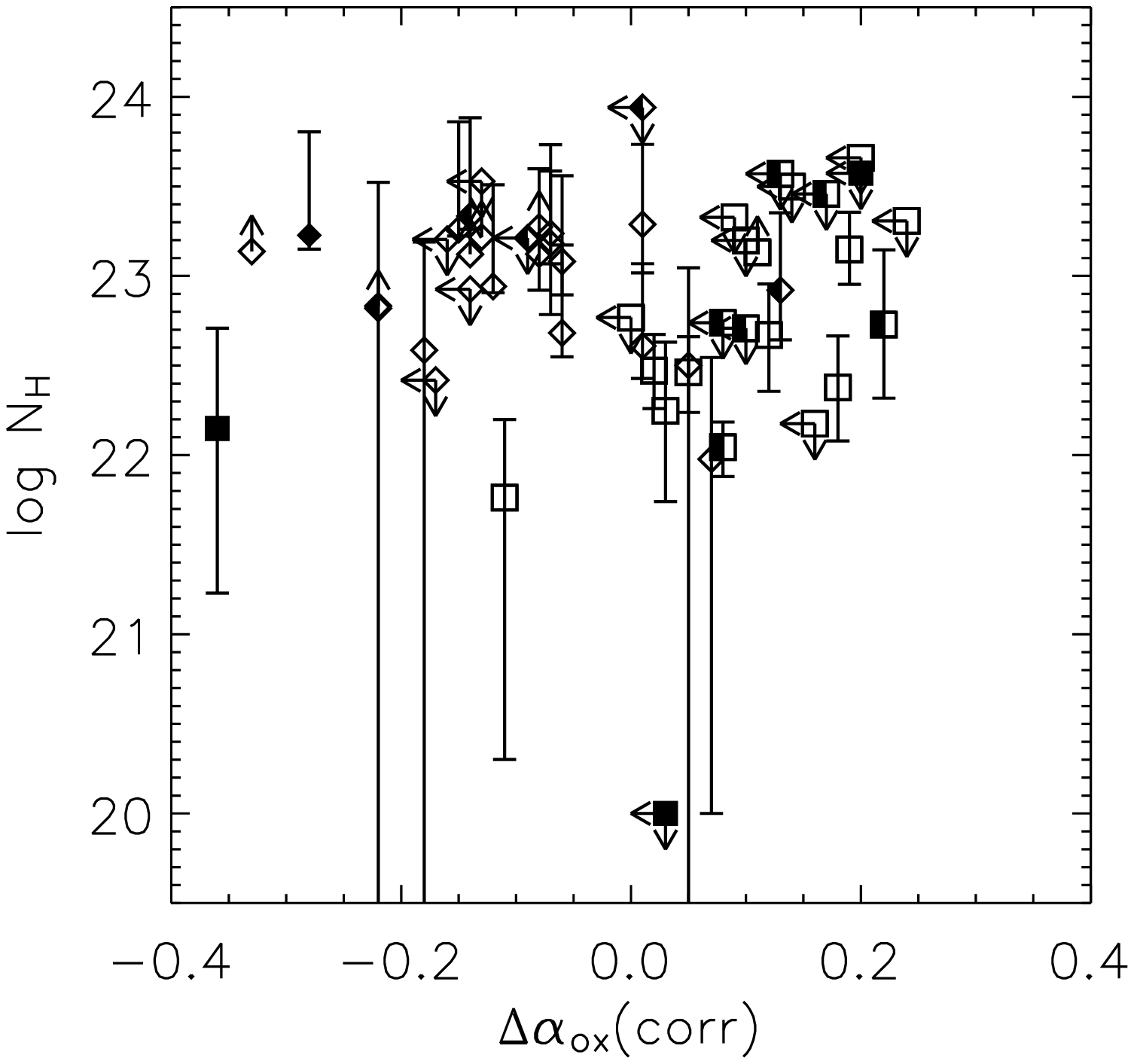}\caption{The left panel shows the \nh\ vs.
\daox\ = \aox\ $-$ \aoxl\ and the right panel shows the \nh\ vs.
\daoxcorr\ = \aoxcorr\ $-$ \aoxl\ for the detected BAL QSOs in the
combined sample.SDSS BAL QSOs in our sample are shown with squares
and LBQS BAL QSOs in G06's sample are shown with diamonds.The
open,filled and half-filled symbols indicate HiBALs,LoBALs and BAL
QSOs of unknown type,respectively. }\label{fig_nhaox}
\end{figure}

First, we check whether \daox\ is a good indicator of X-ray
absorption. In the left panel of the Fig. \ref{fig_nhaox}, we show
\nh\ versus \daox\ for the combined subsample of 51 BAL QSOs that
\nh\ is obtained either from spectral fit or from HR analysis.
Similar to G06, we find a clear correlation between the two
quantities. The probability for null hypothesis is less than 0.01\%~
using non-parametric Kendall $\tau$-test (See Table 4). Then we
compare the redshift distributions of sources with \daox\ $> -0.2$
and \daox\ $< -0.2$.The result is their distributions are very
similar,which indicates that the correlation between \nh\ and \daox\
is not from selection effect of redshift.These suggest that \daox\
can be used as a measure for X-ray absorption indeed.

Next, we examine the correlations between BAL properties and the
X-ray absorption column density \nh\ (Fig. \ref{fig_nhuv})
measured through X-ray spectral fit or HR analysis. We do not find
any correlation with a probability of null hypothesis less than
1\% (Table 4). However, a weak correlation between BI and \nh\
cannot be rejected because the large uncertainty in the \nh\
measurement may reduce the significance of a weak correlation to
the measured level (2\%).

\begin{figure}[t]
\plottwo{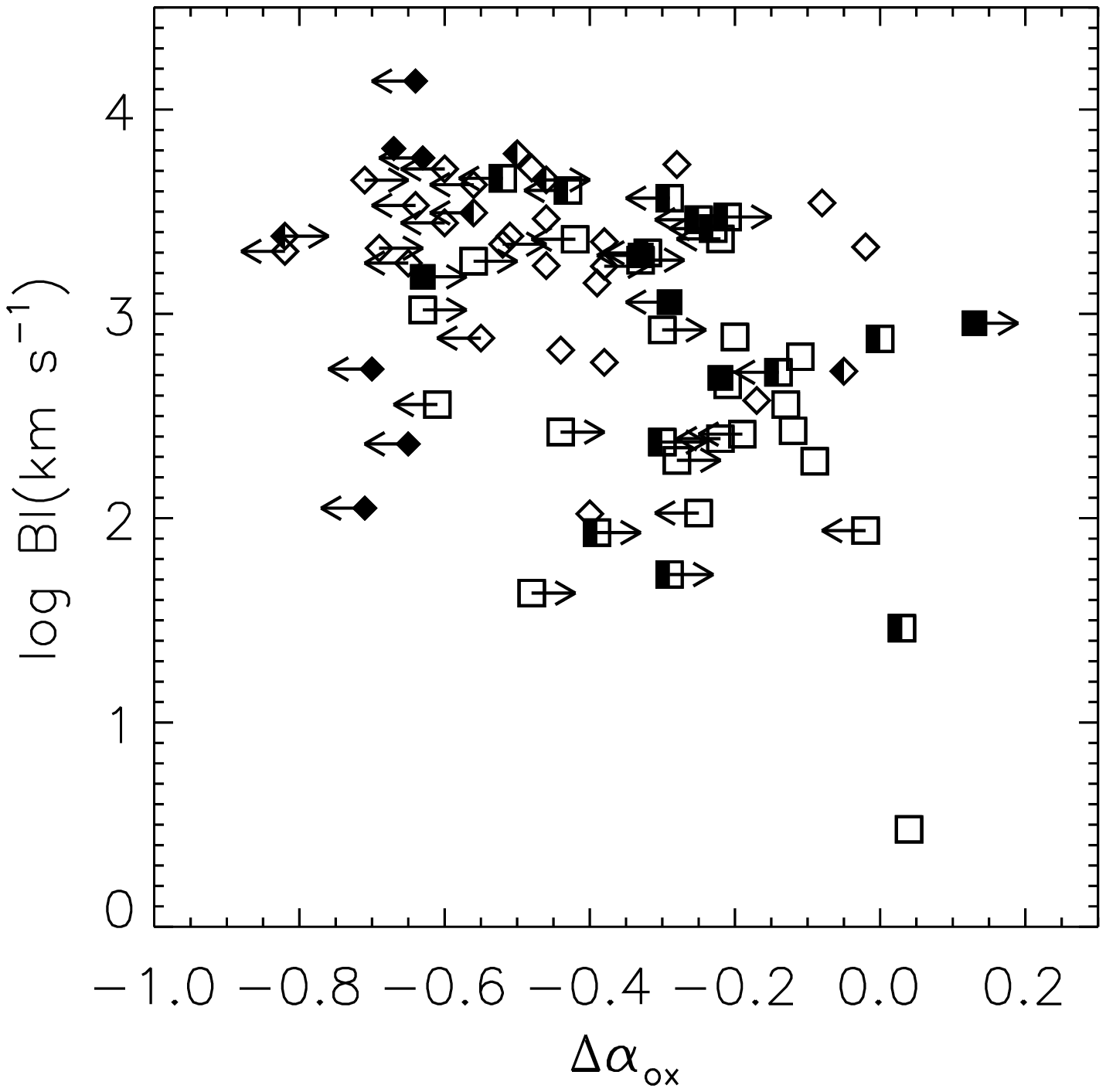}{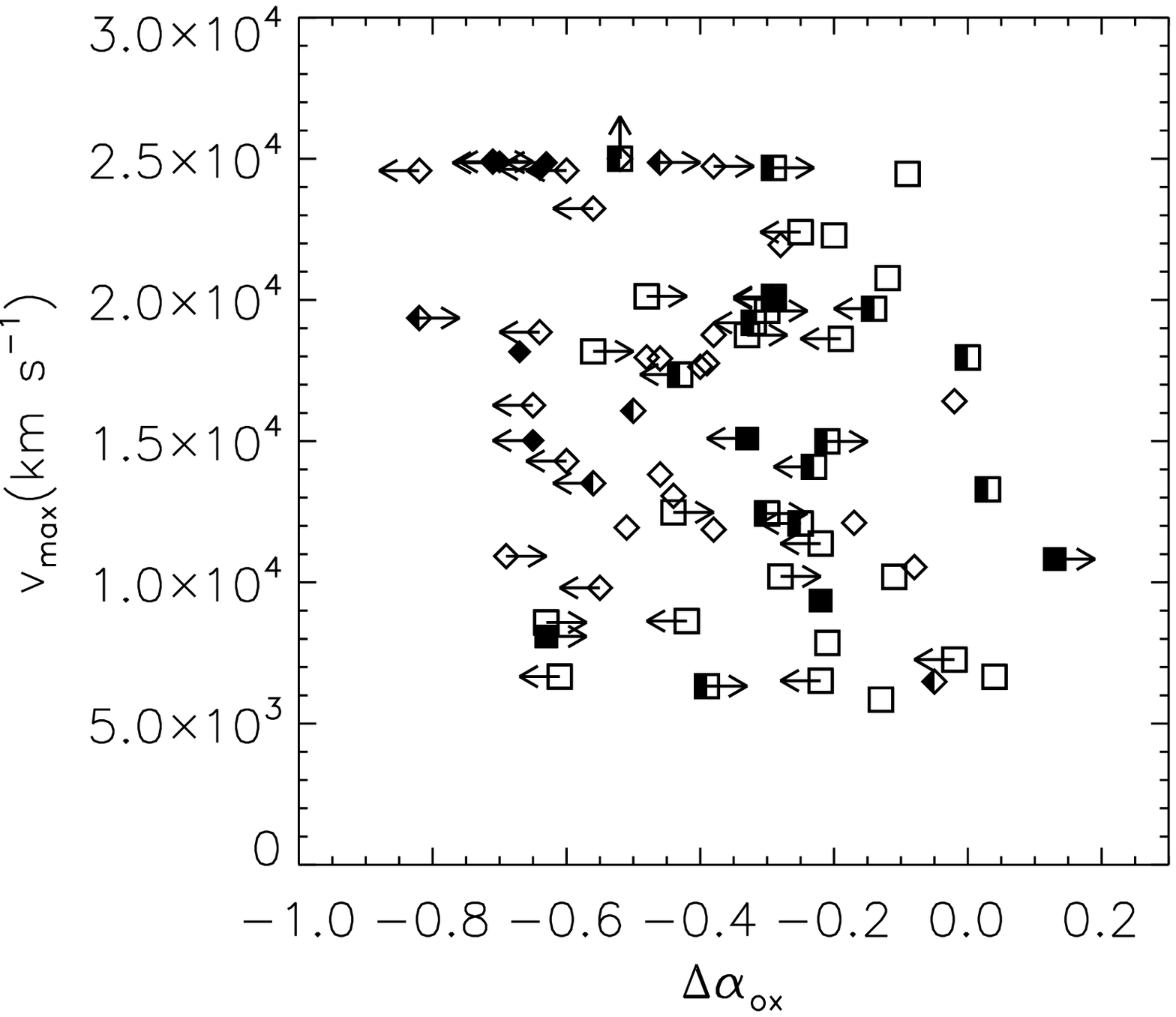}\caption{Plots of \daox\ = \aox\ $-$
\aoxl\ vs. \CIV\ absorption-line parameters for the combined sample
: BALnicity index(BI , left panel) and maximum outflow velocity of
absorption,(\vmax\ , right panel).Symbols are the same as in Fig.
\ref{fig_nhaox} .}\label{fig_daouv}
\end{figure}

\begin{figure}[t]
\plottwo{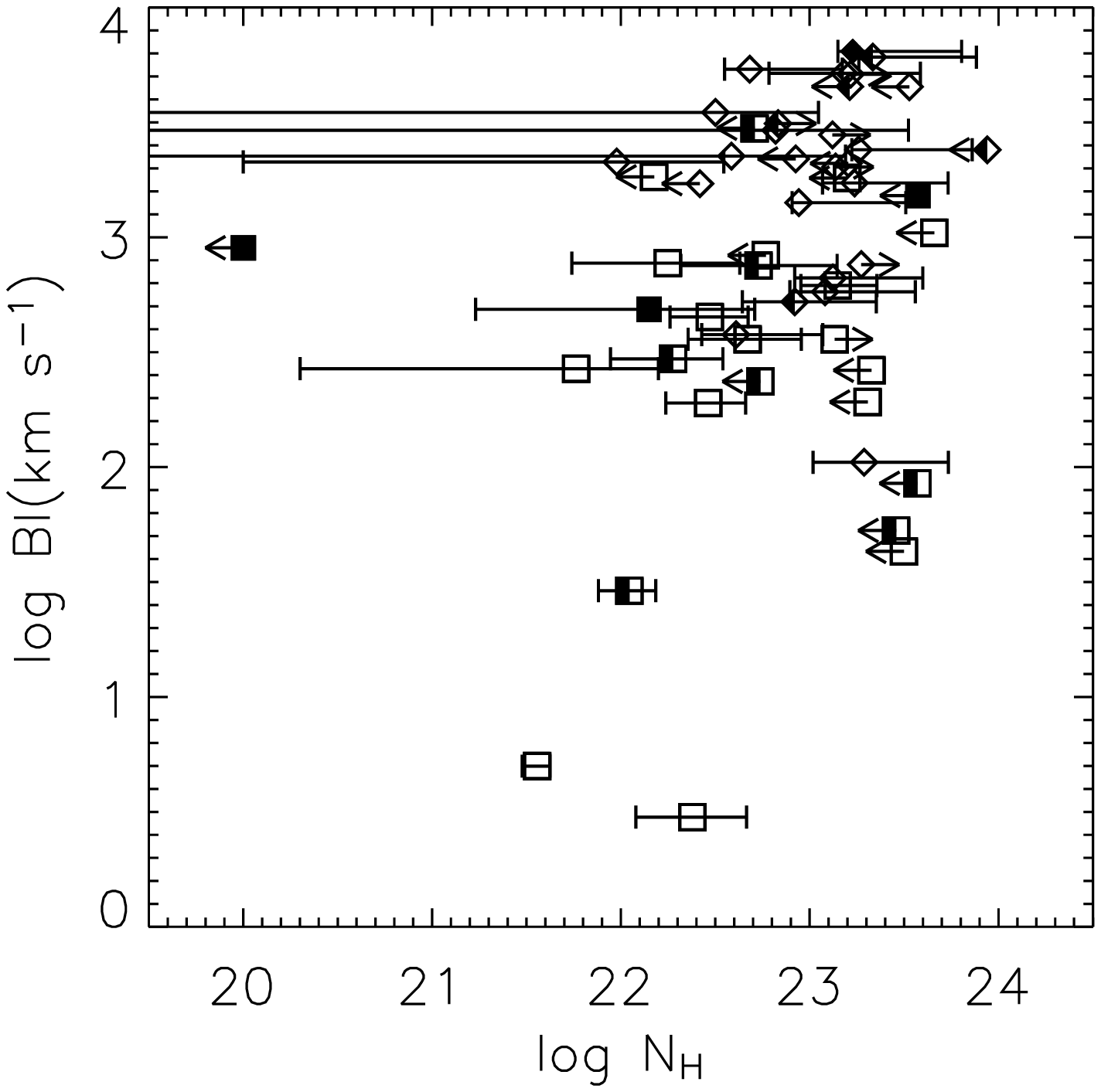}{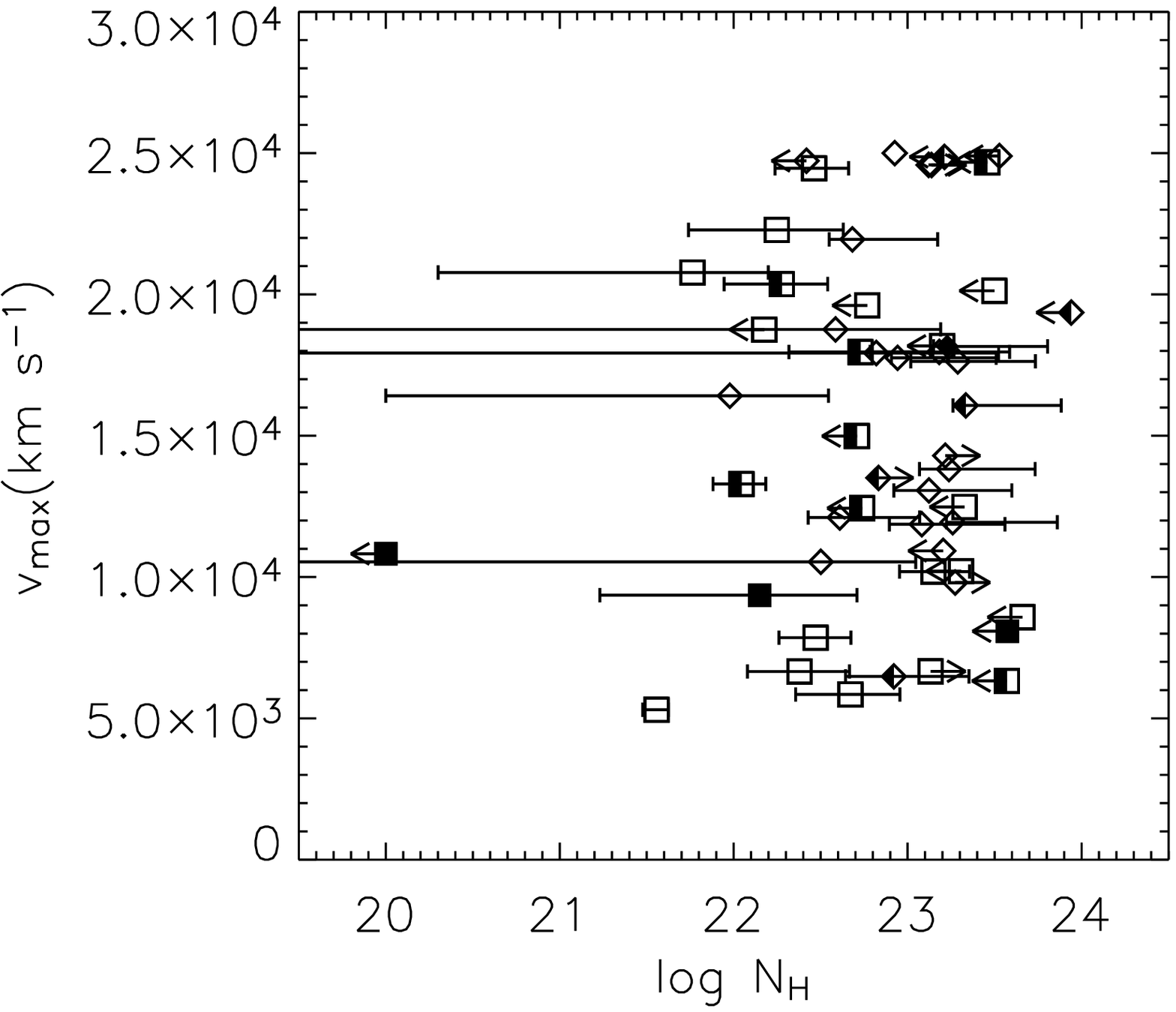}\caption{Plots of \nh\ vs. \CIV\
absorption-line parameters for the detected BAL QSOs in the combined
sample : BALnicity index(BI , left panel) and maximum outflow
velocity of absorption,(\vmax\ , right panel).Symbols are the same
as in Fig. \ref{fig_nhaox} .}\label{fig_nhuv}
\end{figure}

We show \daox\ versus BI and \daox\ versus \vmax\ in Fig.
\ref{fig_daouv}. Two lensed BAL QSOs are excluded from following
analysis because their UV and X-ray light may have been differntly
amplified. There appears a correlation between \daox\ and BI with
the probability for null hypothesis of only 0.05\% (Table 4). The
correlation appears not linear, rather there is an upper envelope.
Since LoBAL QSOs may be different from the HiBAL QSOs
\citep{boroson92,wang07}, we also make Kendall test for 45 HiBAL
QSOs only. The correlation is marginally significant with a null
probability of 1\%. The decrease in significance is caused by
reducing the sample size. However, we do not find any significant
correlation between \vmax\ and \daox, which was seen in G06, in
neither the whole sample nor in the HiBAL subsample with a null
probability of 5\% and 68\%, respectively (Table 4). Comparison with
the LBQS/\chandra\ BAL sample(G06), our sample has a handful BAL
QSOs on the upper right of the figure. These BAL QSOs destroy the
weak correlation trend of \vmax\ vs \daox\ in the LBQS/\chandra\ BAL
sample(G06). These correlations are more or less similar to the
correlations using \nh\ with an exception of higher significance. It
is worthwhile to note that \daox\ reflects a combination of the
X-ray absorption and the intrinsic deviation to the average quasar
SED. Therefore, one must be careful as using it as an indicator of
X-ray absorption. We will discuss below the implication of these
results.

\subsection{Intrinsic X-ray Properties And The Outflow}

Previous studies have shown that UV properties, such as the
blueshift and the equivalent width of \CIV\ emission line, are
correlated with X-ray to optical flux ratio for non-BAL QSOs (e.g.
\citet{baskin04,richards06}). It would be interesting to explore
whether the BAL properties are correlated with the intrinsic \aox.
Unlike the correlation with X-ray absorption, such correlation
should give information for the primary driver of the outflow. Here
we use a corrected \aox\, i.e. \aoxcorr\ to represent the intrinsic
\aox\, and investigate its relation with the UV absorption line
properties. We also study the correlations between UV properties and
\daoxcorr\ so that we can compare the results with those in previous
section and G06. We note that the range of UV luminosity,
$l_{2500}$, for the combined sample is only 2 dex, which introduce a
scatter in \aoxcorr\, through \aox$\sim l_{2500}$ correlation, of
less than 0.27, a factor of about 2.5 smaller than
the dynamic range of \aoxcorr.
Therefore, the difference between \daoxcorr\ and \aoxcorr\ should
be small in correlation analysis. This is verified below
that the relationships between \daoxcorr\ and UV properties have a
very similar behavior as those between \aoxcorr\ and UV properties.

Before exploring UV and X-ray connection, we first check whether
\aoxcorr\ and \daoxcorr\ are affected by the X-ray absorption or
not. We plot \nh\ versus \daoxcorr\ on the right panel of Fig.
\ref{fig_nhaox}. There is no apparent correlation between the two
quantities. Kendall test gives a probability of chance coincidence
of 36\% (Table 4). Therefore, we can conclude that there is no
evidence that \aoxcorr\ and \daoxcorr\ are significantly affected by
X-ray absorption.

\begin{figure}[t]
\plottwo{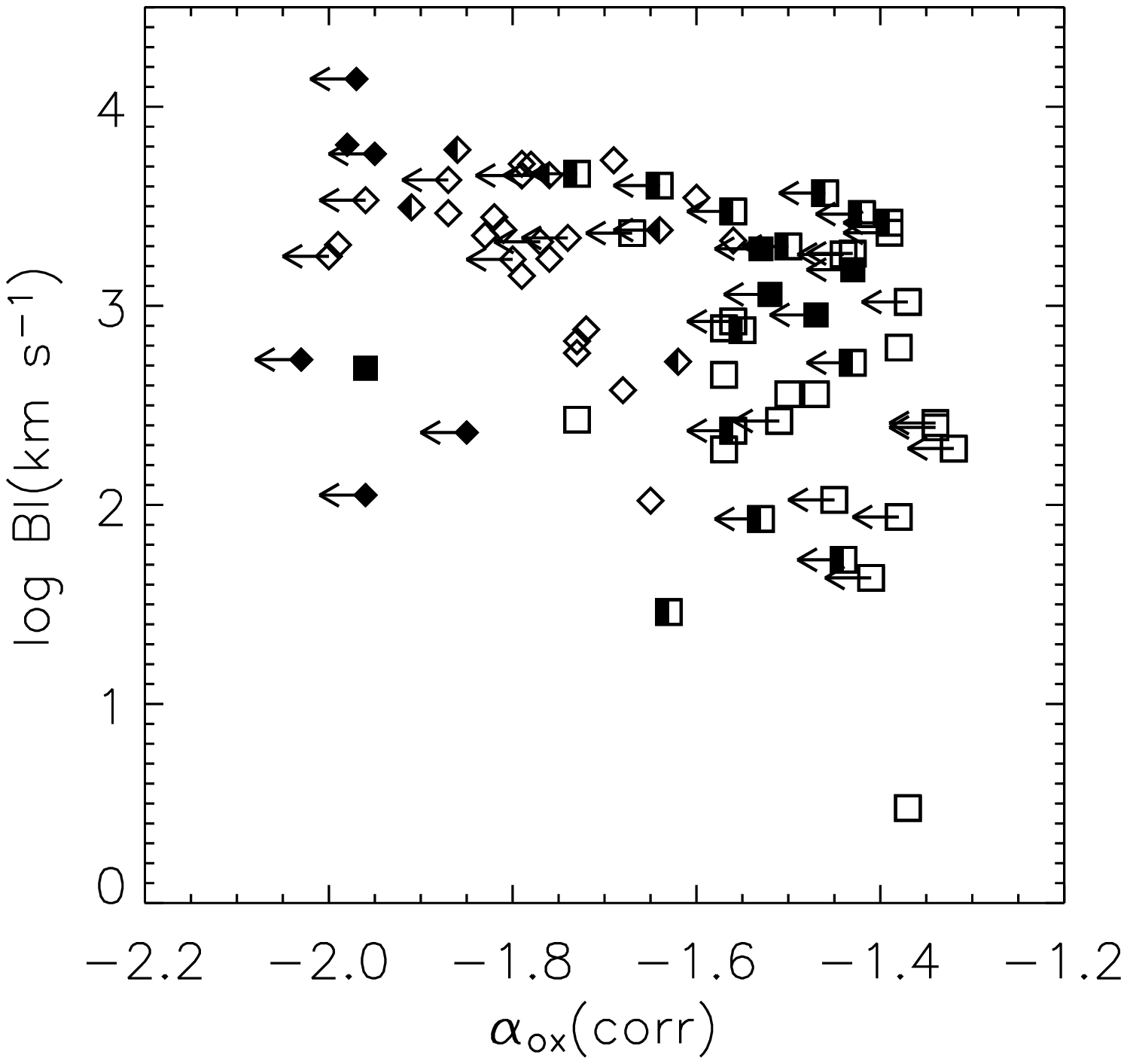}{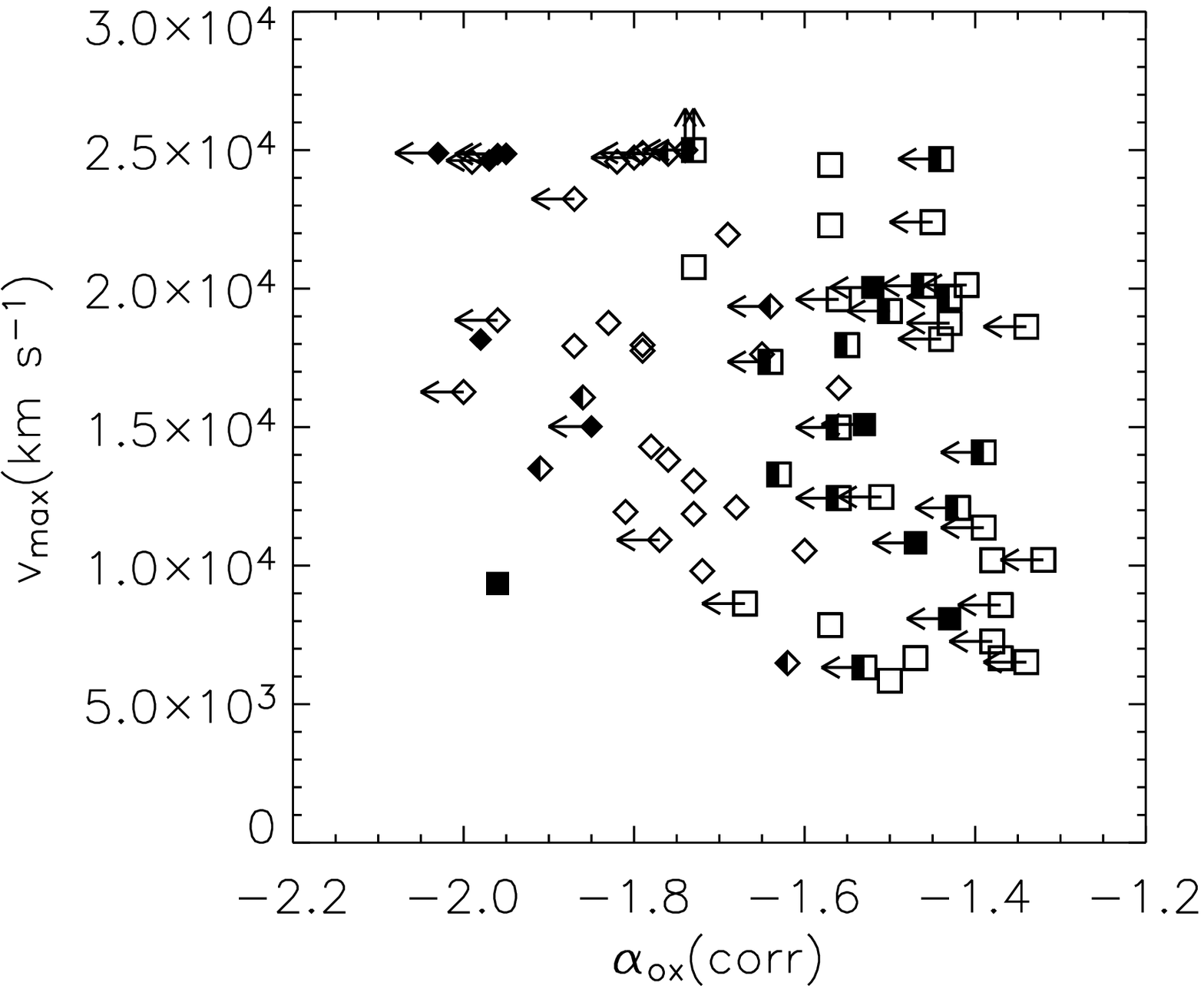} \caption{Plots of \aoxcorr\ vs. \CIV\
absorption-line parameters for the combined sample : BALnicity
index(BI , left panel) and maximum outflow velocity of
absorption,(\vmax\ , right panel).Symbols are the same as in Fig.
\ref{fig_nhaox} .}\label{fig_aocuv}
\end{figure}

\begin{figure}[t]
\plottwo{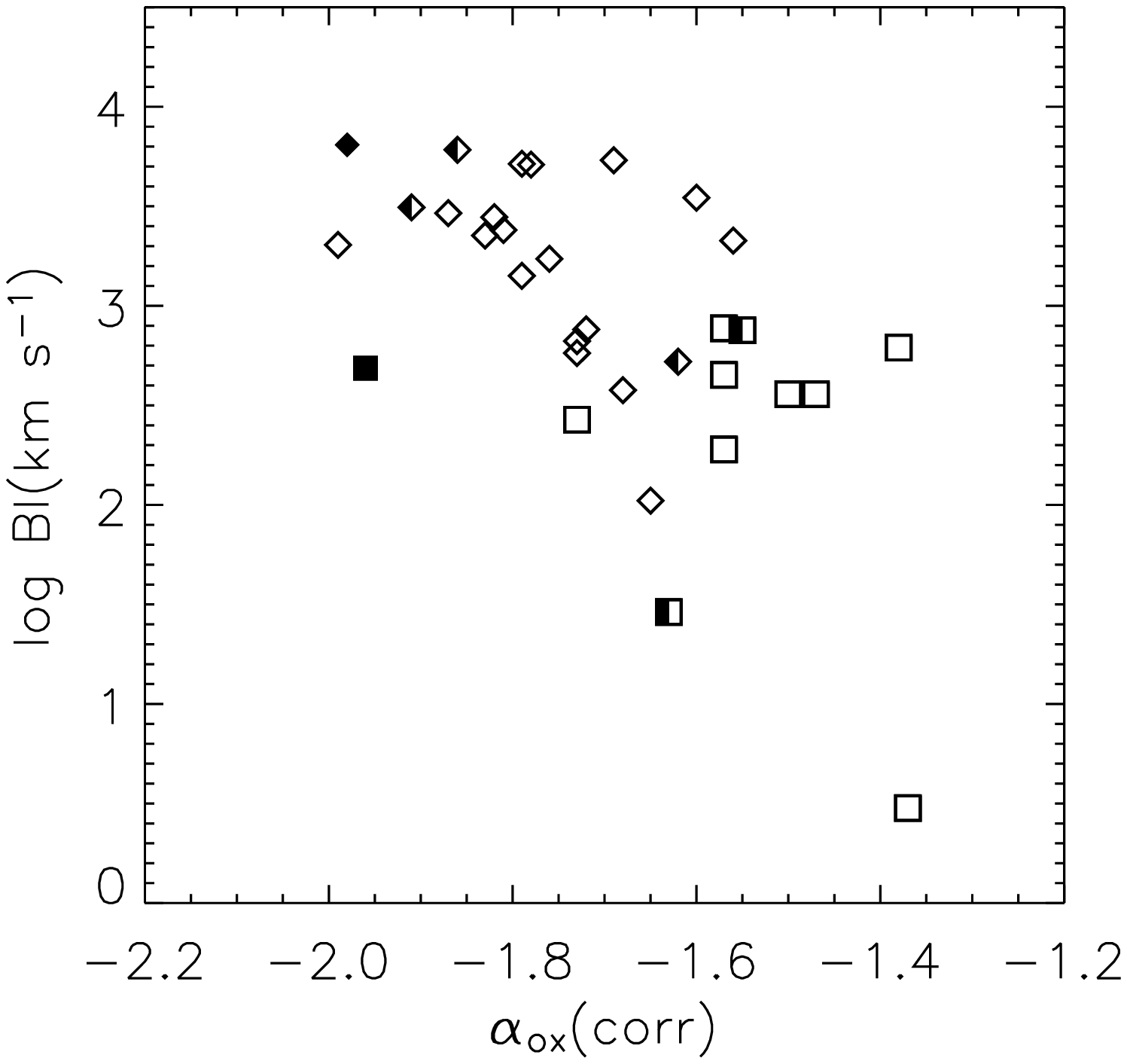}{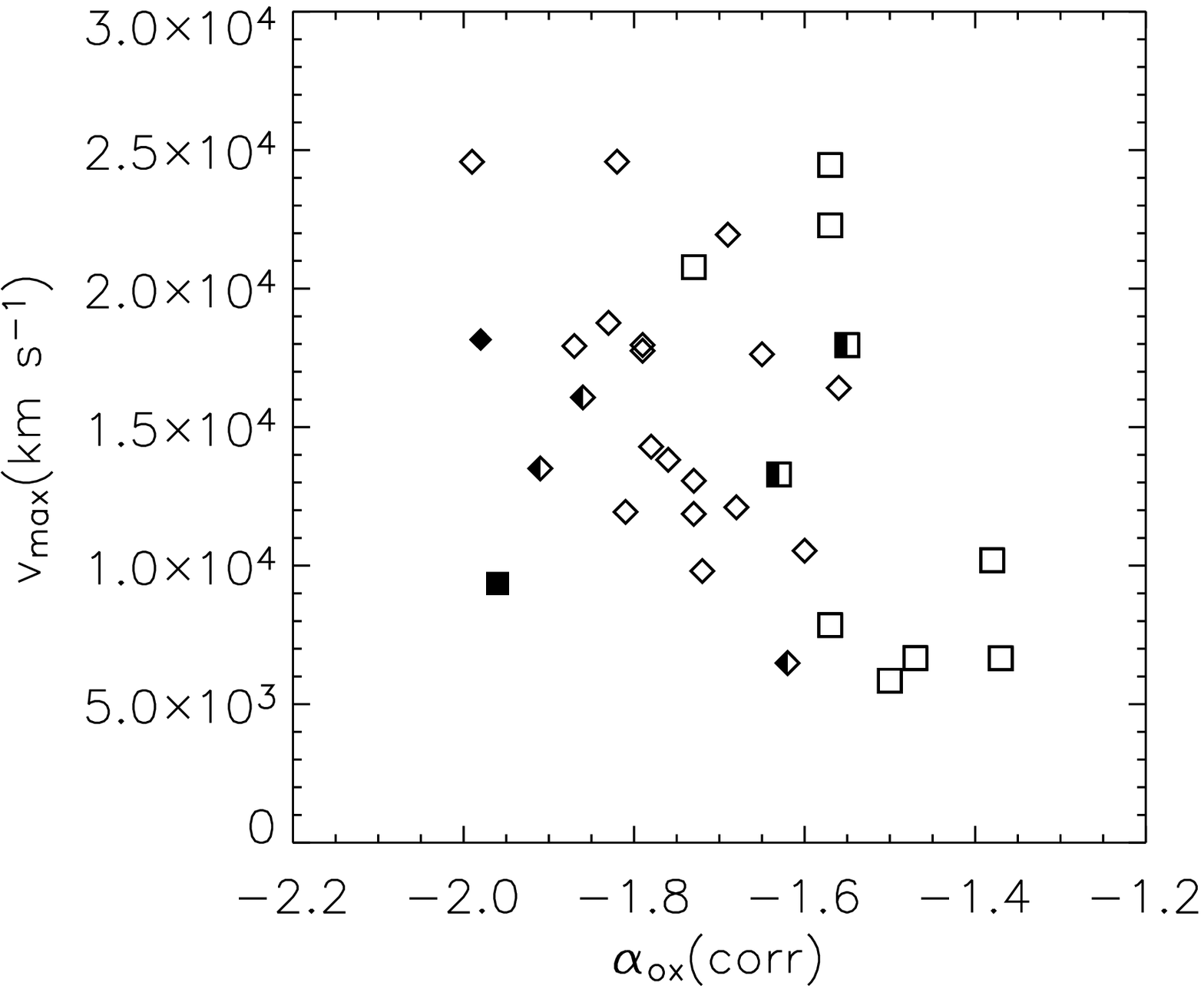}\caption{Plots of \aoxcorr\ vs. \CIV\
absorption-line parameters for the BAL QSOs detected in the hard
X-ray band in the combined sample : BALnicity index(BI , left panel)
and maximum outflow velocity of absorption,(\vmax\ , right
panel).Symbols are the same as in Fig. \ref{fig_nhaox}
.}\label{fig_aocuvd}
\end{figure}

We then explore the correlations between \aoxcorr\ and BI or
\vmax\ with Kendall and Spearman tests. We find that \aoxcorr\ is
significantly correlated with both BI and \vmax\ with a Null
probability of less than 0.1\% for either test(See Fig 5; also
Table 4). The correlation is still very significant ($P<0.1\%$)
for HiBAL QSO subsample (45 QSOs). For clarity we show only QSOs
detected in the hard X-ray band in Fig.\ref{fig_aocuvd}.
Furthermore, these QSOs are more important for the Kendall and
Spearman tests than the rest objects, and can give us a clear
trend about these correlations. We note that one LoBAL QSOs, SDSS
J133553.61+514744.1, which appears largely discrepant with the
main sample in Fig. \ref{fig_aocuv} and \ref{fig_aocuvd}. This
quasar has very steep X-ray photon index $\Gamma\sim2.56$ so that
our `absorption-correction'  underestimate the intrinsic X-ray
luminosity. If we set \aox$=-1.82$ as the lower limit of \aoxcorr
(it is reasonable since \aoxcorr\ is larger than \aox), this
quasar would move rightward and is consistent with other quasars.
The correlations between \daoxcorr\ and UV properties are very
similar to above correlations using \aoxcorr\ except the latter
appear slightly more significant (Table 4).  This may indicate
that the dependence of UV properties on \aoxcorr\ is more
fundamental than on \daoxcorr.

Is it possible that these correlations are introduced by some
selection effect in the sample? If it is the case this effect would
tend to miss the objects which occupy the bottom-left (X-ray weak
and high UV absorption) and top-right (X-ray strong and low UV
absorption) corners of Fig. \ref{fig_aocuv} and \ref{fig_aocuvd}.
Note that the sample selections of ours and G06's are both based on
the optical luminosity and have nothing to do with the X-ray
properties. If the relative X-ray luminosity is uncorrelated with
BAL properties it is hard to understand why G06 and we select the
objects at the bottom-right(top-left) corner but miss those at the
bottom-left(top-right) corner. Since objects at top-right
(bottom-left) corners, if they really exist, should have the same
optical properties as these at top-left (bottom-right). All of these
analysis indicate the correlations between the intrinsic \aox\ and
the properties of UV absorber are real. We discuss the implications
of the correlations in details in the next subsection.

\subsection{Discussion}\label{sec_dis}

Using a larger and more uniform sample, we reexamine the
correlations between BAL properties and X-ray absorption presented
in G06. Two indicators of X-ray absorption, \nh\ and \daox, are used
in the work. We identify the correlation between \daox\ and $BI$ as
the only significant one. In particular, we do not find the
correlation between \daox\ and \vmax\ claimed in G06, and any
correlation between \nh\ and the UV absorption properties. Although
we can not rule out a weak correlation between \nh\ and the UV
absorption line properties due to relative large error bar of \nh\,
our results clearly suggest that X-ray absorption is {\em not} the
major factor that determines UV absorption properties. Given the
fact that almost all BAL QSOs show strong absorption in X-ray, it
seems that X-ray absorption is a necessary condition for launching
of the BAL winds, but the properties of the wind depend on other
factors. As shown above, the observed correlation between \daox\ and
BI may be the secondary effect of the correlation between BI and
\aoxcorr\ or \daoxcorr, as \daox\ is composed of the contributions
of absorption and of \daoxcorr.

In passing, we note that lack of correlations between the X-ray
absorption column density and UV properties does not necessarily
contradict with the scenario of radiatively accelerated wind as
naively thought. For locally optically thin material, the ratio of
the radiation force to the gravitational force is a function of
Eddington ratio and the cross-section ratio of effective absorption
to Thomson scattering. If resonant scattering is responsible for the
absorption opacity, the cross-section will be determined by the
fraction Li-like ions. According to the equatorial wind
model\citep{murray95}, a clump of highly ionized gas (shielding
gas), which accounts for most X-ray opacity, blocks the soft X-ray
interior to the wind. The transmitted flux of soft X-rays that
ionize Li-like ions in the wind depends strongly on the X-ray column
density, \nh. If \nh\ along the direction is very small,
high-velocity wind cannot be launched because of the reduction of
the radiation force caused by the over-ionization. On the other
hand, if \nh\ is very large, the wind will end up with a turbulent
flow due to the blocking of thick very-low-ionized gas
behind\citep[see the figure 4 of ][]{proga00}. Thus, high-velocity
wind can only be launched when the radial column density \nh\ is
moderate as the fraction of Li-like ions, such as \CIV\ and \NV, is
large enough. As far as the X-ray absorption column density is in
the right range, the fraction of Li-like ions should be the dominant
species. The flow properties are then determined self-consistently
by the launching radius, the gas density at the launching radius and
the radiation intensity. If X-ray absorber is well separated from
the UV absorber, then we would not expect any correlation between
the X-ray absorption column density and the flow properties for BAL
QSOs apart.

On the other hand, the X-ray absorber may
be the 'hitchhiking' gas just located at
the inner edge of the wind, and its properties may have a close
connection with the boundary conditions of the disk wind\citep{murray95}.
In that case, we should consider globally the structure of gas along
a line of sight. The wind starts at a radius where the radiation force
is substantially larger than the gravitational force. As far as the gas
density is high enough, such a region can certainly exist. \citet{murray95}
has worked out a consistent line acceleration model, and they found
that gas column density and final velocity are correlated for a constant
Eddington ratio and at a given launch radius. However, if the launching
radius is not exactly scaled with luminosity as $L^{1/2}$ and there are a
range of Eddington ratio, as they assumed, the correlation can be smeared
out.

More interesting results of our work are the strong correlations
between the parameters of outflow and intrinsic \aox. We argue that
these correlations are essential rather than due to some selection
effect or the secondary effect of other correlations (see previous
subsection for details). In fact our results are consistent with
\citet{richards06} who found that the QSOs with large blueshifts of
\CIV\ emission line, i.e. the parent population of BAL QSOs as
suggested by \citet{richards06}, tend to have lower X-ray luminosity
for given optical luminosity (their figure 5). It is also upheld by
\citet{laor02} who presented significant correlation between the
equivalent width of \CIV\ absorption and \aox\ in a sample of
non-BAL QSOs. This correlation is actually predicted by
\citet{murray95}, in which they found that quasars with a large
X-ray to UV ratio can only produce weak low velocity winds while
quasars with a small X-ray to UV ratio can produce strong and large
velocity winds. This is exactly what we have found here. As we
discussed above, their model also predicted a correlation between
the X-ray absorption column density and the maximum velocity of the
flow, which is not observed in this sample. Lack of such correlation
may be due to two important factors that (1) variation in the
Eddington ratio and launching radius; (2) the large uncertainties in
the measurement of absorption column density.

We do not fully understand why the variation in the Eddington
ratio and launching does not completely smeared out the correlation
with \aoxcorr. There seems one reason for this.
\citet{wang04} found that the 2-10kev luminosities
to bolometric luminosities ratio tightly anti-correlated with
Eddington ratio for a sample of broad-line and narrow-line Seyfert 1
AGNs. If this correlation holds up for BAL QSOs, one would expect
that quasars with high Eddington ratio would have larger radiative
acceleration force, or large terminal velocity, and at the same time
X-ray weaker. \citet{ganguly07} find that \vmax\ as a function of
Eddington ratio has an upper envelope in the SDSS3 BAL catalog,
exactly as expected. Since there is no clear correlation of BI with
UV luminosities\citep[cf.][]{laor02}, Eddington rate and black hole
mass\citep{ganguly07},it is very likely that the BAL properties
are more likely determined by the SED of quasars rather than Eddington ratio.

\section{SUMMARY}\label{sec_sum}

We compile a large \CIV\ BAL QSOs sample from the \xmm\ archive data
and SDSS DR5. The sample consists of 41 BAL QSOs, among which 26
QSOs are detected in the X-ray band. Our sample spans wide and
homogeneous ranges of both BI and \vmax\ and can be used to
complement the LBQS/\chandra\ BAL sample(G06). In addition, the
combined sample of ours and G06s show a more homogeneous
distribution of intrinsic X-ray properties than G06s. Using this
combined sample, we investigate the correlations between X-ray and
UV properties of BAL QSOs.

We briefly summarize our conclusions below:
\begin{enumerate}
\item{We confirm the previous results that BAL QSOs are generally
soft X-ray weak, which is mainly due to the intrinsic X-ray
absorption. We also find the X-ray luminosities of BAL QSOs with
given optical luminosity have large scatter. The scatter is
 caused by both the various column densities of X-ray absorber and
 the scatter of intrinsic X-ray emission at given optical luminosity.}
\item{We do not find any evidence for the claimed correlation between the BAL
properties and soft X-ray absorption, with an exception of the
correlation between BI and \daox. The correlation between BI and
\daox\ can be induced by the correlation between BI and the
intrinsic \aox. The X-ray absorber is important for launching the
high-velocity wind but do not directly determine the BAL
properties.}
\item{There are significant correlations between intrinsic X-ray strength,
\aoxcorr, and BI and \vmax\ in the combined sample. These
correlations are essential rather than due to any artificiality. We
preliminarily interpret that the BAL properties are influenced by
the intrinsic SED of QSOs, which is consistent with the prediction
of a radiatively accelerated disk wind model \citep{murray95}.}
\end{enumerate}

\acknowledgements We thank the anonymous referee for helpful comments.We thank
Paul Hewett for providing electronic data of LBQS spectra.We also acknowledge
the Sloan Digital Sky Survey(http://www.sdss.org).
This work was supported by the Knowledge
Innovation Program of the Chinese Academy of Sciences, Grant No.
KJCX2-YW-T05 and the National Basic Research Program of China (973 Program)
under Grant No. 2007CB815400.

\clearpage
\input{tab1}
\clearpage
\input{tab2}
\clearpage
\input{tab3}
\clearpage
\clearpage
\input{tab4}
\clearpage

\end{document}

%% file: tab1.tex
\begin{deluxetable}{cccrrrrcc}
\rotate \tabletypesize{\scriptsize}\tablecaption{Observed BAL QSOs
\label{tab1}} \tablewidth{0pt} \tablehead{
\colhead{Name(SDSS)\tablenotemark{a}}& \colhead{z\tablenotemark{b}}
& \colhead{i\tablenotemark{c}} &
\colhead{$f_{2500}$\tablenotemark{d}} & \colhead{\nh\
\tablenotemark{e}} & \colhead{BI\tablenotemark{f}} & \colhead{\vmax\
\tablenotemark{g}}
&  \colhead{BAL Type\tablenotemark{h}} & \colhead{$R_i$\tablenotemark{i}}     \\
} \startdata
 J020230.66-075341.2 & 1.722 & 18.69 &  4.69 & 2.13 &  245 &  6518 & Hi & $<0.68$ \\
 J023224.87-071910.5 & 1.597 & 18.14 &  9.81 & 3.02 &  450 &  7855 & Hi & $<0.44$ \\
 J024304.68+000005.4 & 1.995 & 18.19 &  7.96 & 3.56 &  360 &  5847 & Hi & $<0.81$ \\
 J085551.24+375752.2 & 1.929 & 18.19 &  8.98 & 2.92 &  268 & 20777 & Hi & $<0.40$ \\
 J090928.50+541925.9 & 3.760 & 19.58 &  1.02 & 2.09 & 2890 & 12086 &  H & $<0.97$ \\
 J091127.61+055054.1 & 2.793 & 17.77 & 11.38 & 3.64 &  296 & 20368 &  H & $<0.32$ \\
 J091400.95+410600.9 & 2.052 & 19.51 &  2.10 & 1.80 &  258 & 18625 & Hi & $<0.92$ \\
 J092138.45+301546.9 & 1.590 & 17.97 &  9.77 & 1.90 &  264 & 12479 & Hi & $<0.41$ \\
 J092238.43+512121.2 & 1.753 & 20.11 &  1.36 & 1.43 &  900 & 10823 & Lo & $<1.14$ \\
 J092345.19+512710.0 & 2.168 & 19.14 &  3.18 & 1.42 & 1832 & 18755 & Hi & 1.42    \\
 J092507.54+521102.6 & 2.995 & 19.10 &  4.22 & 1.46 & 2982 & 14990 &  H & $<0.80$ \\
 J094309.56+481140.5 & 1.809 & 18.72 &  5.43 & 1.21 &  106 & 22401 & Hi & $<0.59$ \\
 J094440.42+041055.6 & 1.984 & 18.23 &  7.42 & 3.63 & 1934 & 15095 & Lo & $<0.50$ \\
 J095110.56+393243.9 & 1.716 & 19.68 &  1.97 & 1.57 & 2328 & 11371 & Hi & $<1.07$ \\
 J100728.69+534326.7 & 1.772 & 19.06 &  3.37 & 0.74 &    3 &  6656 & Hi & $<0.80$ \\
 J105201.35+441419.8 & 1.791 & 18.47 &  5.06 & 1.12 &  617 & 10192 & Hi & $<0.55$ \\
 J110853.98+522337.9 & 1.665 & 18.53 &  5.81 & 0.90 &  87  &  7266 & Hi & $<0.55$ \\
 J111816.95+074558.1 & 1.735 & 15.89 & 64.89 & 3.53 &    5 &  5306 & Hi & $<-0.45$\\
 J112055.78+431412.5 & 2.389 & 18.66 &  4.22 & 2.07 & 3684 & 20103 &  H & $<0.64$ \\
 J112432.14+385104.3 & 3.530 & 19.99 &  1.75 & 2.06 &  236 & 12438 &  H & $<1.24$ \\
 J113419.96+485805.7 & 3.080 & 20.05 &  2.29 & 1.60 & 1988 & 19192 &  H & $<1.26$ \\
 J113406.87+525959.0 & 1.769 & 18.84 &  4.71 & 1.11 & 1046 &  8581 & Hi & $<0.67$ \\
 J120449.77+020635.6 & 2.776 & 19.24 &  3.43 & 1.88 &  518 & 19690 &  H & $<0.85$ \\
 J120522.18+443140.4 & 1.921 & 18.42 &  5.62 & 1.27 &  772 & 22284 & Hi & $<0.57$ \\
 J122708.29+012638.4 & 1.954 & 19.07 &  2.94 & 1.84 &  835 & 19612 & Hi & $<1.58$ \\
 J125741.41+565214.2 & 1.841 & 19.36 &  2.59 & 1.27 & 1811 & 18178 & Hi & $<1.00$ \\
 J132827.07+581836.9 & 3.140 & 18.53 &  3.49 & 1.37 & 85   &  6323 &  H & $<0.27$ \\
 J133004.72+472301.0 & 2.825 & 19.16 &  2.38 & 1.55 & 4020 & 17357 &  H & 1.18    \\
 J133553.61+514744.1 & 1.838 & 18.08 &  7.48 & 1.11 &  486 &  9359 & Lo & 1.14    \\
 J133639.40+514605.2 & 2.229 & 19.04 &  2.65 & 1.11 & 2319 &  8631 & Hi & $<0.80$ \\
 J134145.12-003631.0 & 2.215 & 18.52 &  5.86 & 2.06 & 1519 &  8085 & Lo & $<0.58$ \\
 J142555.22+373900.7 & 2.731 & 19.13 &  2.69 & 0.94 &   53 & 24676 &  H & $<0.55$ \\
 J142539.38+375736.7 & 1.897 & 18.02 &  8.52 & 0.95 &  190 & 24464 & Hi & $<0.08$ \\
 J142652.94+375359.9 & 1.812 & 19.12 &  3.43 & 0.95 &  43  & 20131 & Hi & $<0.57$ \\
 J144027.00+032637.9 & 2.136 & 18.74 &  4.05 & 2.77 & 1141 & 20041 & Lo & $<0.68$ \\
 J144625.48+025548.6 & 1.883 & 18.97 &  4.31 & 3.06 &  360 &  6665 & Hi & $<0.71$ \\
 J150824.22-000603.8 & 1.578 & 18.44 &  6.73 & 4.58 &  192 & 10208 & Hi & $<0.51$ \\
 J152553.89+513649.1 & 2.883 & 16.57 & 29.85 & 1.57 &  754 & 17965 &  H & $<-0.46$\\
 J153229.97+323658.4 & 3.048 & 19.22 &  1.94 & 2.03 & 2614 & 14090 &  H & $<0.85$ \\
 J154359.44+535903.2 & 2.370 & 16.96 & 19.76 & 1.25 &   29 & 13292 &  H & $<-0.11$\\
 J164151.84+385434.2 & 3.776 & 18.51 &  7.56 & 1.21 & 4610 & $>25000$ &  H & $<0.57$ \\
\enddata
\tablenotetext{a}{SDSS
ID.}\tablenotetext{b}{redshift.}\tablenotetext{c}{i band fiber
magnitude of SDSS.}\tablenotetext{d}{The rest-frame 2500\AA\ flux
density(in units of $10^{-17}$ \flambda).} \tablenotetext{e}{The
values for \nh\ (in units of 10$^{20}$\cmsq\ ) are from Galactic HI\
maps \citep{dickey90}.} \tablenotetext{f}{The BALnicity Index (BI;
in units of \kms\ )} \tablenotetext{g}{The maximum outflow
velocity(\vmax; in units of \kms\ )} \tablenotetext{h}{The BAL
subtype.``Hi" denotes a HiBAL-only object;``Lo" denotes a LoBAL
detected through Mg II absorption;``H" denotes a HiBAL object in
which the Mg II region is not within the spectral coverage.}
\tablenotetext{i}{The radio-to-optical flux ratios,
$R_{i}=log(S_{1.4GHz}/S_{i})$, following the definition of
Ivezi{\'c} et al. (2002).}
\end{deluxetable}

%% file: tab2.tex
\begin{deluxetable}{clcrrcccc}
\tabletypesize{\scriptsize} \rotate \tablecaption{\xmm\ Observing
Log\label{tab2}} \tablewidth{0pt} \tablehead{ \colhead{Name(SDSS)} &
\colhead{Obs.ID\tablenotemark{a}}  &  \colhead{Date}  &
\colhead{$T_{exp}$\tablenotemark{b}} & \colhead{Instrument}  &
\colhead{Soft\tablenotemark{c}}
  &  \colhead{Hard\tablenotemark{c}}   &  \colhead{Counts Rate\tablenotemark{d}}       &  \colhead{HR\tablenotemark{e}}  \\
} \startdata

J020230.66-075341.2  &  0411980201  &  2006-07-03  &   8.16  &  pn  &  $<7.9$                &  $<16.5$               &  $<1.81$                 &    ... \\
J023224.87-071910.5  &  0200730401  &  2004-01-07  &  37.51  &  pn  &  $64^{+17.6}_{-15.8}$  &  $35^{+15.6}_{-13.8}$  &  $2.64^{+0.62}_{-0.57}$  &  $-0.29^{+0.17}_{-0.19}$ \\
J024304.68+000005.4  &  0111200101  &  2000-07-29  &  38.51  &  mos1&  $76^{+19.3}_{-17.5}$  &  $54^{+17.0}_{-15.2}$  &  $3.39^{+0.66}_{-0.61}$  &  $-0.17^{+0.13}_{-0.12}$ \\
J085551.24+375752.2  &  0302581801  &  2005-10-10  &  28.45  &  mos1&  $48^{+16.4}_{-14.7}$  &  $12^{+12.9}_{-11.5}$  &  $2.11^{+0.77}_{-0.71}$  &  $-0.60^{+0.05}_{-0.18}$ \\
J090928.50+541925.9  &  0200960101  &  2005-03-28  &  71.05  &  mos2&  $<18.0$               &  $<22.3$               &  $<0.40$                 &    ... \\
J091127.61+055054.1  &  0083240201(T)  &  2001-11-02  &   8.93  &  pn  &  $277^{+35.9}_{-34.2}$ &  $151^{+31.1}_{-29.2}$ &  $47.94^{+5.24}_{-5.10}$ &  $-0.29^{+0.08}_{-0.08}$ \\
J091400.95+410600.9  &  0147671001  &  2003-04-27  &  13.37  &  mos1&  $<5.8$                &  $<8.6$                &  $<0.72$                 &    ... \\
J092138.45+301546.9  &  0150620101  &  2003-04-23  &  15.76  &  mos1&  $11^{+8.7}_{-6.8}$    &  $<23.4$               &  $1.59^{+0.79}_{-0.68}$  &$<0.35$ \\
J092238.43+512121.2  &  0300910301  &  2005-10-08  &  15.57  &  pn  &  $173^{+27.5}_{-25.7}$ &  $<13.8$               &  $10.73^{+2.02}_{-1.91}$ &$<-0.85$\\
J092345.19+512710.0  &  0300910301  &  2005-10-08  &  15.55  &  pn  &  $31^{+17.3}_{-15.6}$  &  $<9.8$                &  $0.32^{+1.54}_{-0.31}$  &$<-0.52$\\
J092507.54+521102.6  &  0201130501  &  2004-11-15  &  46.54  &  mos1&  $45^{+13.8}_{-12.0}$  &  $<29.5$               &  $1.35^{+0.38}_{-0.34}$  &$<-0.21$\\
J094309.56+481140.5  &  0201470101  &  2004-10-14  &  30.35  &  mos1&  $<14.6$               &  $<25.2$               &  $<0.95$                 &    ... \\
J094440.42+041055.6  &  0201290301  &  2004-05-18  &  21.27  &  mos1&  $<6.5$                &  $<11.1$               &  $<0.50$                 &    ... \\
J095110.56+393243.9  &  0111290101  &  2001-11-03  &  21.10  &  mos1&  $<11.1$               &  $<11.7$               &  $<0.77$                 &    ... \\
J100728.69+534326.7  &  0070340201  &  2001-05-10  &  19.69  &  pn  &  $135^{+24.1}_{-22.3}$ &  $43^{+17.3}_{-15.5}$  &  $9.04^{+1.49}_{-1.40}$  &   $-0.52^{+0.11}_{-0.12}$\\
J105201.35+441419.8  &  0146990901  &  2003-05-24  &   5.11  &  pn  &  $32^{+12.2}_{-10.4}$  &  $25^{+13.1}_{-11.2}$  &  $11.16^{+3.42}_{-3.06}$ &   $-0.12^{+0.23}_{-0.25}$\\
J110853.98+522337.9  &  0304071201  &  2005-10-21  &   8.18  &  mos2&  $<21.3$               &  $<8.1$                &  $<3.09$                 &    ... \\
J111816.95+074558.1  &  0203560401(T)  &  2004-06-26  &  68.34  &  pn  &$8762^{+162.2}_{-161.8}$&  $2598^{+93.9}_{-92.1}$&  $166.24^{+2.79}_{-2.69}$&   $-0.54^{+0.01}_{-0.01}$\\
J112055.78+431412.5  &  0107860201  &  2001-05-08  &  21.96  &  mos1&  $<5.8$                &  $<13.4$               &  $<0.57$                 &    ... \\
J112432.14+385104.3  &  0052140201  &  2001-12-03  &  24.55  &  pn  &  $31^{+14.8}_{-13.0}$  &  $<20.4$               &  $1.63^{+0.80}_{-0.72}$  &$<-0.21$\\
J113419.96+485805.7  &  0149900201  &  2003-11-24  &  14.96  &  pn  &  $<13.6$               &  $<17.5$               &  $<1.43$                 &    ... \\
J113406.87+525959.0  &  0200431301  &  2004-11-04  &  10.82  &  mos2&  $2^{+6.9}_{-1.8}$     &  $<10.4$               &  $0.52^{+0.80}_{-0.51}$  & $<0.70$\\
J120449.77+020635.6  &  0093060101  &  2001-12-21  &  14.21  &  mos1&  $<11.6$               &  $<7.4$                &  $<0.99$                 &    ... \\
J120522.18+443140.4  &  0156360101  &  2003-06-11  &  23.95  &  pn  &  $92^{+22.2}_{-20.3}$  &  $47^{+21.0}_{-19.2}$  &  $5.80^{+1.26}_{-1.18}$  &  $-0.32^{+0.17}_{-0.20}$ \\
J122708.29+012638.4  &  0110990201  &  2001-06-23  &   9.56  &  pn  &  $22^{+14.0}_{-12.2}$  &  $<12.4$               &  $2.51^{+1.79}_{-1.60}$  &$<-0.28$\\
J125741.41+565214.2  &  0081340201  &  2001-06-07  &  21.39  &  mos1&  $4^{+6.5}_{-4.0}$     &  $<6.1$                &  $0.12^{+0.42}_{-0.12}$  & $<0.18$\\
J132827.07+581836.9  &  0405690201  &  2006-11-19  &  25.97  &  pn  &  $27^{+18.5}_{-16.7}$  &  $<27.6$               &  $1.42^{+1.03}_{-0.97}$  & $<0.00$\\
J133004.72+472301.0  &  0112840201  &  2003-01-15  &  17.11  &  pn  &  $<12.3$               &  $<7.9$                &  $<0.69$                 &    ... \\
J133553.61+514744.1  &  0084190201  &  2002-06-12  &  38.39  &  pn  &  $86^{+26.0}_{-24.3}$  &  $6^{+21.2}_{-5.9}$    &  $2.40^{+0.90}_{-0.84}$  & $-0.87^{+0.11}_{-0.13}$  \\
J133639.40+514605.2  &  0084190201  &  2002-06-12  &  37.24  &  pn  &  $<44.7$               &  $<10.2$               &  $<0.78$                 &    ... \\
J134145.12-003631.0  &  0111281601  &  2002-07-20  &   7.41  &  mos1&  $2^{+5.3}_{-2.0}$     &  $<5.0$                &  $0.23^{+0.90}_{-0.23}$  & $<0.40$\\
J142555.22+373900.7  &  0112230201  &  2002-12-18  &  19.48  &  pn  &  $29^{+13.0}_{-11.2}$  &  $<28.6$               &  $2.34^{+0.89}_{-0.80}$  &$<-0.01$\\
J142539.38+375736.7  &  0112230201  &  2002-12-18  &  19.48  &  pn  &  $109^{+24.2}_{-22.5}$ &  $36^{+19.4}_{-17.6}$  &  $7.44^{+1.57}_{-1.49}$  & $-0.50^{+0.15}_{-0.20}$  \\
J142652.94+375359.9  &  0112230201  &  2002-12-18  &  19.48  &  pn  &  $14^{+14.5}_{-13.2}$  &  $<32.9$               &  $1.61^{+1.15}_{-1.05}$  & $<0.41$\\
J144027.00+032637.9  &  0300210701  &  2006-01-08  &  23.05  &  mos1&  $<19.5$               &  $<17.7$               &  $<1.08$                 &   ...  \\
J144625.48+025548.6  &  0203050801  &  2005-01-12  &   7.86  &  mos1&  $<3.1$                &  $4^{+5.4}_{-3.9}$     &  $0.13^{+0.82}_{-0.13}$  & $>0.13$\\
J150824.22-000603.8  &  0305750201  &  2005-07-20  &   5.03  &  mos2&  $6^{+6.3}_{-4.5}$     &  $<11.3$               &  $2.18^{+1.65}_{-1.29}$  & $<0.34$\\
J152553.89+513649.1  &  0011830401(T)  &  2001-12-13  &   2.82  &  pn  &  $122^{+24.7}_{-22.9}$ &  $56^{+18.6}_{-16.9}$  & $63.05^{+10.86}_{-10.17}$& $-0.37^{+0.12}_{-0.13}$  \\
J153229.97+323658.4  &  0039140101  &  2002-07-30  &   4.74  &  pn  &  $<8.7$                &  $<11.7$               &  $<2.97$                 &  ...   \\
J154359.44+535903.2  &  0060370901(T)  &  2002-02-06  &  16.18  &  pn  &  $571^{+43.1}_{-41.5}$ &  $115^{+24.8}_{-22.9}$ &  $42.40^{+3.06}_{-2.93}$ &  $-0.66^{+0.04}_{-0.05}$ \\
J164151.84+385434.2  &  0204340101  &  2004-08-20  &  12.22  &  pn  &  $<7.7$                &  $<17.0$               &  $<1.27$                 &   ...  \\

\enddata
\tablenotetext{a}{(T) means the object is the intended PI target of
the \xmm\ observation.} \tablenotetext{b}{The effective exposure
time in $10^3\ s$.} \tablenotetext{c}{Errors are 1$\sigma$ Poisson
errors\citep{gehr86} for detections,and for non-detections the
limits are the $90\%$ confidence limits from Bayesian
statistics\citep{kraft91}.The count rate is the full energy
band,0.3-10.0keV.} \tablenotetext{d}{The count rate in $10^{-3}\
counts\ s^{-1}$} \tablenotetext{e}{$HR$ is defined as
$(h-s)/(h+s)$.The $HR$ errors are propagated from the counts errors
using the Bayesian estimation of \citet{park06}.Notice the
difference of the definitions of the hard and soft band between
ours($s:0.3-2.0$~keV;$h:2.0-10.0$~keV) and
G06s($s:0.5-2.0$~keV;$h:2.0-8.0$~keV).  }

\end{deluxetable}

%% file: tab3.tex
\begin{deluxetable}{crrrrrrrrrrrr}
\rotate
\tabletypesize{\scriptsize} \tablecaption{X-RAY
PROPERTIES\label{tab3}} \tablewidth{0pt} \tablehead{
\colhead{Name(SDSS)} & \colhead{$N_H$\tablenotemark{a}}  &
\colhead{$\Gamma_{HR}$\tablenotemark{b}} &
\colhead{$log(f_x)$\tablenotemark{c}} &
\colhead{$log(f_{2kev})$\tablenotemark{d}} &
\colhead{$log(f_{2500})$\tablenotemark{d}} &
\colhead{$log(l_{2500})$\tablenotemark{e}}  &
\colhead{$\alpha_{ox}$} &
\colhead{$\Delta\alpha_{ox}$\tablenotemark{f}} &
\colhead{$\alpha_{ox}(corr)$\tablenotemark{g}}
 & \colhead{$\Delta\alpha_{ox}(corr)$\tablenotemark{h}}  \\

} \startdata
 J020230.66$-$075341.2 & $\cdots$ & $\cdots$ & $<-13.418$ & $<-31.782$ & $-27.140$ & 30.726 & $<-1.78$ & $<-0.22$ & $<-1.34$ &  $<0.22$ \\
 J023224.87$-$071910.5 &    $2.96^{+1.76}_{-1.14}$ & $1.06^{+0.32}_{-0.26}$ &  $-13.230$ &  $-31.547$ & $-26.861$ & 30.945 &  $-1.80$ &  $-0.21$ &  $-1.57$ &     0.02 \\
 J024304.68+000005.4 &    $4.69^{+4.34}_{-2.42}$ & $1.12^{+0.21}_{-0.22}$ &  $-13.095$ &  $-31.376$ & $-26.827$ & 31.155 &  $-1.75$ &  $-0.13$ &  $-1.50$ &     0.12 \\
 J085551.24+375752.2 &    $0.58^{+1.00}_{-0.56}$ & $1.93^{+0.52}_{-0.12}$ &  $-13.576$ &  $-31.331$ & $-26.794$ & 31.162 &  $-1.74$ &  $-0.12$ &  $-1.73$ &   $-0.11$\\
 J090928.50+541925.9 & $\cdots$ &  $\cdots$ & $<-13.814$ & $<-32.178$ & $-27.316$ & 31.129 & $<-1.87$ & $<-0.25$ & $<-1.42$ &  $<0.20$ \\
 J091127.61+055054.1 &    $1.89^{+1.58}_{-1.01}$ & $1.22^{+0.13}_{-0.11}$ &  $-12.620$ &  $-30.787$ & $-26.467$ & 31.768 &  $-1.66$ &  $\cdots$ &  $-1.41$ &  $\cdots$ \\
 J091400.95+410600.9 & $\cdots$ & $\cdots$  & $<-13.537$ & $<-31.902$ & $-27.390$ & 30.613 & $<-1.73$ & $<-0.19$ & $<-1.34$ &  $<0.21$ \\
 J092138.45+301546.9 & $<21.25$&  $>0.21$ &  $-13.137$ &  $-32.144$ & $-26.864$ & 30.939 &  $>-2.03$ &  $>-0.44$ & $<-1.51$ &  $<0.09$ \\
 J092238.43+512121.2 & $<0.01$ &  $>2.38$ &  $-13.560$ &  $-31.205$ & $-27.669$ & 30.212 &  $>-1.36$ &  $ >0.13$ & $<-1.47$ &  $<0.03$ \\
 J092345.19+512710.0 & $<1.50$ &  $>1.47$ &  $-14.155$ &  $-32.167$ & $-27.177$ & 30.869 &  $>-1.92$ &  $>-0.33$ & $<-1.43$ &  $<0.16$ \\
 J092507.54+521102.6 & $<5.11$ &  $>1.12$ &  $-13.463$ &  $-31.721$ & $-26.852$ & 31.433 &  $>-1.87$ &  $>-0.21$ & $<-1.56$ &  $<0.10$ \\
 J094309.56+481140.5 & $\cdots$ &  $\cdots$ & $<-13.436$ & $<-31.803$ & $-27.049$ & 30.856 & $<-1.83$ & $<-0.25$ & $<-1.45$ &  $<0.14$ \\
 J094440.42+041055.6 & $\cdots$&   $\cdots$ & $<-13.571$ & $<-31.931$ & $-26.861$ & 31.116 & $<-1.95$ & $<-0.33$ & $<-1.53$ &  $<0.09$ \\
 J095110.56+393243.9 & $\cdots$ &  $\cdots$ & $<-13.651$ & $<-32.017$ & $-27.518$ & 30.346 & $<-1.73$ & $<-0.22$ & $<-1.39$ &  $<0.13$ \\
 J100728.69+534326.7 &    $2.39^{+2.24}_{-1.19}$ & $1.42^{+0.24}_{-0.19}$ &  $-13.131$ &  $-31.206$ & $-27.268$ & 30.621 &  $-1.51$ &     0.04 &  $-1.37$ &     0.18 \\
 J105201.35+441419.8 &   $14.07^{+8.61}_{-5.09}$ & $0.89^{+0.37}_{-0.33}$ &  $-13.012$ &  $-31.464$ & $-27.086$ & 30.811 &  $-1.68$ &  $-0.11$ &  $-1.38$ &     0.19 \\
 J110853.98+522337.9 & $\cdots$ & $\cdots$ & $<-12.834$ & $<-31.201$ & $-27.065$ & 30.775 & $<-1.59$ & $<-0.02$ & $<-1.38$ &  $<0.19$ \\
 J111816.95+074558.1 &    $0.36^{+0.06}_{-0.06}$ & $1.64^{+0.02}_{-0.02}$ &  $-12.228$ &  $-30.155$ & $-25.995$ & 31.877 &  $-1.60$ &   $\cdots$ &  $-1.51$ & $\cdots$ \\
 J112055.78+431412.5 & $\cdots$& $\cdots$ & $<-13.599$ & $<-31.963$ & $-26.995$ & 31.124 & $<-1.91$ & $<-0.29$ & $<-1.46$ &  $<0.15$ \\
 J112432.14+385104.3 & $<5.48$ &  $>1.03$ &  $-13.840$ &  $-32.175$ & $-27.126$ & 31.276 &  $>-1.94$ &  $>-0.30$ & $<-1.56$ &  $<0.08$ \\
 J113419.96+485805.7 & $\cdots$&  $\cdots$ & $<-13.816$ & $<-32.182$ & $-27.100$ & 31.205 & $<-1.95$ & $<-0.32$ & $<-1.50$ &  $<0.13$ \\
 J113406.87+525959.0 & $<45.73$& $>-0.48$ &  $-13.158$ &  $-32.835$ & $-27.123$ & 30.765 &  $>-2.19$ &  $>-0.63$ & $<-1.37$ &  $<0.20$ \\
 J120449.77+020635.6 & $\cdots$ & $\cdots$ & $<-13.251$ & $<-31.616$ & $-26.992$ & 31.238 & $<-1.78$ & $<-0.14$ & $<-1.43$ &  $<0.20$ \\
 J120522.18+443140.4 &    $1.77^{+2.50}_{-1.22}$ & $1.16^{+0.33}_{-0.24}$ &  $-13.422$ &  $-31.666$ & $-27.000$ & 30.952 &  $-1.79$ &  $-0.20$ &  $-1.57$ &     0.03 \\
 J122708.29+012638.4 & $<5.86$ &  $>1.14$ &  $-13.842$ &  $-32.099$ & $-27.272$ & 30.694 &  $>-1.85$ &  $>-0.30$ & $<-1.56$ &  $<0.00$ \\
 J125741.41+565214.2 & $<15.81$&  $>0.40$ &  $-13.959$ &  $-32.827$ & $-27.361$ & 30.558 &  $>-2.10$ &  $>-0.56$ & $<-1.44$ &  $<0.10$ \\
 J132827.07+581836.9 & $<37.22$&  $>0.67$ &  $-13.540$ &  $-32.228$ & $-26.904$ & 31.415 &  $>-2.04$ &  $>-0.39$ & $<-1.53$ &  $<0.13$ \\
 J133004.72+472301.0 & $\cdots$&  $\cdots$& $<-14.098$ & $<-32.464$ & $-27.140$ & 31.103 & $<-2.04$ & $<-0.43$ & $<-1.64$ & $<-0.02$ \\
 J133553.61+514744.1 &    $1.41^{+3.70}_{-1.24}$ & $2.56^{+2.04}_{-0.45}$ &  $-14.048$ &  $-31.642$ & $-26.901$ & 31.017 &  $-1.82$ &  $-0.22$ &  $-1.96$ &   $-0.36$\\
 J133639.40+514605.2 & $\cdots$& $\cdots$ & $<-14.084$ & $<-32.451$ & $-27.239$ & 30.828 & $<-2.00$ & $<-0.42$ & $<-1.67$ & $<-0.09$ \\
 J134145.12$-$003631.0 & $<37.45$& $>-0.01$ &  $-13.475$ &  $-32.773$ & $-26.899$ & 31.163 &  $>-2.26$ &  $>-0.63$ & $<-1.43$ &  $<0.20$ \\
 J142555.22+373900.7 & $<28.69$&  $>0.70$ &  $-13.413$ &  $-32.058$ & $-27.108$ & 31.110 &  $>-1.90$ &  $>-0.29$ & $<-1.44$ &  $<0.17$ \\
 J142539.38+375736.7 &    $2.90^{+1.68}_{-1.17}$ & $1.49^{+0.46}_{-0.26}$ &  $-13.246$ &  $-31.267$ & $-26.826$ & 31.117 &  $-1.71$ &  $-0.09$ &  $-1.57$ &     0.05 \\
 J142652.94+375359.9 & $<31.55$&  $>0.04$ &  $-13.358$ &  $-32.551$ & $-27.248$ & 30.659 &  $>-2.04$ &  $>-0.48$ & $<-1.41$ &  $<0.14$ \\
 J144027.00+032637.9 & $\cdots$ &  $\cdots$ & $<-13.617$ & $<-31.980$ & $-27.081$ & 30.953 & $<-1.88$ & $<-0.29$ & $<-1.52$ &  $<0.07$ \\
 J144625.48+025548.6 & $>13.56$&  $<0.56$ &  $-14.079$ &  $-32.807$ & $-27.126$ & 30.811 &  $<-2.18$ &  $<-0.61$ &  $-1.47$ &     0.11 \\
 J150824.22$-$000603.8 & $<20.29$&  $>0.24$ &  $-12.869$ &  $-31.846$ & $-27.030$ & 30.767 &  $>-1.85$ &  $>-0.28$ & $<-1.32$ &  $<0.24$ \\
 J152553.89+513649.1 &    $5.36^{+8.61}_{-3.28}$ & $1.33^{+0.22}_{-0.19}$ &  $-12.556$ &  $-30.636$ & $-26.028$ & 32.230 &  $-1.77$ &     0.00 &  $-1.55$ &     0.22 \\
 J153229.97+323658.4 & $\cdots$ &  $\cdots$ & $<-13.615$ & $<-31.980$ & $-27.179$ & 31.119 & $<-1.84$ & $<-0.23$ & $<-1.39$ &  $<0.22$ \\
 J154359.44+535903.2 &    $1.11^{+0.42}_{-0.35}$ & $1.79^{+0.12}_{-0.08}$ &  $-12.889$ &  $-30.684$ & $-26.330$ & 31.783 &  $-1.67$ &     0.03 &  $-1.63$ &     0.08 \\
 J164151.84+385434.2 & $\cdots$&  $\cdots$ & $<-13.961$ & $<-32.327$ & $-26.444$ & 32.004 & $<-2.26$ & $<-0.52$ & $<-1.73$ &  $<0.00$ \\

\enddata
\tablenotetext{a}{Intrinsic absorption column density in units of
$10^{22} cm^{-2}$,its value or upper or lower limit is determined by
fitting the observed spectrum,or comparing the observed HR to a
simulated HR that takes the instrument response into account,or the
upper limit of count rates,assuming $\Gamma=2.0$ and a simple
neutral absorption.} \tablenotetext{b}{$\Gamma_{HR}$ , following the
definition of G06s,is a coarse measure of the hardness of the X-ray
spectrum determined by comparing the observed HR to a simulated HR
that takes the instrument response into account.}
\tablenotetext{c}{The full-band X-ray flux, $f_{\rm x}$, has units
of \flux.} \tablenotetext{d}{X-ray and optical flux densities were
measured at rest-frame 2~keV and 2500\,\AA, respectively; units are
\fnu.} \tablenotetext{e}{The 2500 \AA\ monochromatic
luminosity,$l_{\rm 2500}$, has units of \lumin~Hz$^{-1}$.  The
redshift bandpass correction has been included.}
\tablenotetext{f}{The parameter \daox\ is the difference between the
  observed \aox\ and \aoxl, the predicted \aox\ from $l_{\rm 2500}$
  calculated from Equation 6 of \citep{stra05}.}
\tablenotetext{g}{The parameter \aoxcorr\  is \aox\ calculated
  assuming $\Gamma=2.0$ and using the hard-band count rate to
  normalize the X-ray continuum.}
\tablenotetext{h}{\daoxcorr=\aoxcorr$-$\aoxl.}

\end{deluxetable}

%% file: tab4.tex
\begin{deluxetable}{lrrrr}
\tablecolumns{5} \tablewidth{0pt} \tablecaption{Results from
Non-Parametric Bivariate Statistical Tests \label{tab:stats}}
\tablehead{ \colhead{Variables\tablenotemark{a}} &
\multicolumn{2}{r}{Generalized Kendall} &
\multicolumn{2}{r}{Spearman} \\
\colhead{Independent/Dependent} & \colhead{$\tau$}        &
\colhead{Prob.\tablenotemark{b}(\%)}        & \colhead{$\rho$} &
\colhead{Prob.\tablenotemark{b}(\%)}
} \startdata
\nh/\daox(51)                     &  6.930   &$<0.01$ & $\cdots$    & $\cdots$ \\
\nh/\daoxcorr(51)                 &  0.913   & 36.2   & $\cdots$    & $\cdots$ \\
\nh/BI (53)                       &  2.296   &  2.2   & $\cdots$    & $\cdots$ \\
\nh/\vmax(53)                     &  0.365   & 71.5   & $\cdots$    & $\cdots$ \\
\\
\daox/BI(74)                      &  3.467   & 0.05   & $\cdots$    & $\cdots$ \\
\daox/\vmax(74)                   &  2.003   &  4.5   & $\cdots$    & $\cdots$ \\
\daox/BI(HiBALs only  45)         &  2.553   &  1.1   & $\cdots$    & $\cdots$ \\
\daox/\vmax(HiBALs only  45)      &  0.414   & 67.9   & $\cdots$    & $\cdots$ \\
\\
\aoxcorr/BI(74)                   &  3.657   & 0.03   & $-0.384$    & 0.1  \\
\aoxcorr/\vmax(74)                &  3.936   & 0.01   & $-0.458$    & 0.01 \\
\aoxcorr/BI(HiBALs only 45)       &  3.447   & 0.06   & $-0.492$    & 0.1  \\
\aoxcorr/\vmax(HiBALs only 45)    &  3.335   & 0.09   & $-0.469$    & 0.2  \\
\\
\daoxcorr/BI(74)                  &  3.215   & 0.1    & $-0.329$    &  0.5  \\
\daoxcorr/\vmax(74)               &  3.623   & 0.03   & $-0.425$    &  0.03 \\
\daoxcorr/BI(HiBALs only 45)      &  3.071   & 0.2    & $-0.430$    &  0.4 \\
\daoxcorr/\vmax(HiBALs only 45)   &  3.218   & 0.1    & $-0.473$    &  0.2 \\
\enddata
\tablenotetext{a}{The number of data points is given in
parentheses.} \tablenotetext{b}{The probability(in units of percent)
that the given variables are not correlated. Spearman's $\rho$
cannot be calculated for data with both upper and lower limits.}
\end{deluxetable}